\documentclass[aps,prd,twocolumn,preprintnumbers,nofootinbib,amsmath,amssymb,superscriptaddress,10pt]{revtex4-2}

\usepackage{graphicx}
\usepackage{dcolumn}
\usepackage{bm}
\usepackage{amsmath}
\usepackage{braket}
\usepackage{slashed}    
\usepackage{epstopdf}
\usepackage{placeins}
\usepackage[shortlabels]{enumitem}
\usepackage{xcolor}
\usepackage{multirow}
\usepackage{makecell}
\usepackage{slashed}
\usepackage{hyperref}
\usepackage{cleveref}
\usepackage[normalem]{ulem}
\usepackage[font=small, labelfont=bf, textfont={small,it}]{caption}

\newcommand{\beq}{\begin{eqnarray}}
\newcommand{\eeq}{\end{eqnarray}}
\newcommand{\beqnn}{\begin{eqnarray*}}
\newcommand{\eeqnn}{\end{eqnarray*}}

\newcommand{\ls}{\ell_{\scriptscriptstyle{7}}}
\newcommand{\lt}{\bar{\ell}_{\scriptscriptstyle{3}}}
\newcommand{\lf}{\bar{\ell}_{\scriptscriptstyle{4}}}
\newcommand{\Nf}{N_{\scriptscriptstyle{\rm f}}}
\newcommand{\Ns}{N_{\scriptscriptstyle{\rm s}}}
\newcommand{\ChPT}{\chi\mathrm{PT}}
\newcommand{\mf}{m_{\scriptscriptstyle{\rm f}}}
\newcommand{\mqu}{m_{\scriptscriptstyle{\rm u}}}
\newcommand{\mqd}{m_{\scriptscriptstyle{\rm d}}}
\newcommand{\mqs}{m_{\scriptscriptstyle{\rm s}}}
\newcommand{\mql}{m_{\scriptscriptstyle{\ell}}}
\newcommand{\LQCDsub}{{\scriptscriptstyle{\rm LQCD}}}
\newcommand{\QCDsub}{{\scriptscriptstyle{\rm QCD}}}
\newcommand{\isoQCDsub}{{\scriptscriptstyle{\rm isoQCD}}}
\newcommand{\YMsub}{{\scriptscriptstyle{\rm YM}}}
\newcommand{\IBsub}{{\scriptscriptstyle{\rm IB}}}
\newcommand{\Lsup}{{\scriptscriptstyle{(\mathrm{L})}}}
\newcommand{\stagsup}{{\scriptscriptstyle{(\mathrm{stag})}}}
\newcommand{\stagsub}{{\scriptscriptstyle{\mathrm{stag}}}}
\newcommand{\physsup}{{\scriptscriptstyle{(\mathrm{phys})}}}
\newcommand{\lsub}{{\scriptscriptstyle{\ell}}}
\newcommand{\usub}{{\scriptscriptstyle{\rm u}}}
\newcommand{\dsub}{{\scriptscriptstyle{\rm d}}}
\newcommand{\connsup}{{\scriptscriptstyle{(\rm conn)}}}
\newcommand{\discsup}{{\scriptscriptstyle{(\rm disc)}}}
\newcommand{\isoQCDsup}{{\scriptscriptstyle{(\rm isoQCD)}}}
\newcommand{\effsup}{{\scriptscriptstyle{(\rm eff)}}}
\newcommand{\effcsup}{{\scriptscriptstyle{(\rm eff,c)}}}
\newcommand{\effdsup}{{\scriptscriptstyle{(\rm eff,d)}}}
\newcommand{\fourptsub}{{\scriptscriptstyle{4\mathrm{pt}}}}
\newcommand{\Ssup}{{\scriptscriptstyle{(\rm S)}}}
\newcommand{\Ssub}{{\scriptscriptstyle{\rm S}}}
\newcommand{\Asub}{{\scriptscriptstyle{\rm A}}}
\newcommand{\Psub}{{\scriptscriptstyle{\rm P}}}
\newcommand{\Vsub}{{\scriptscriptstyle{\rm V}}}

\newcommand{\monec}{\raisebox{-0.4\totalheight}{\includegraphics[scale=0.075]{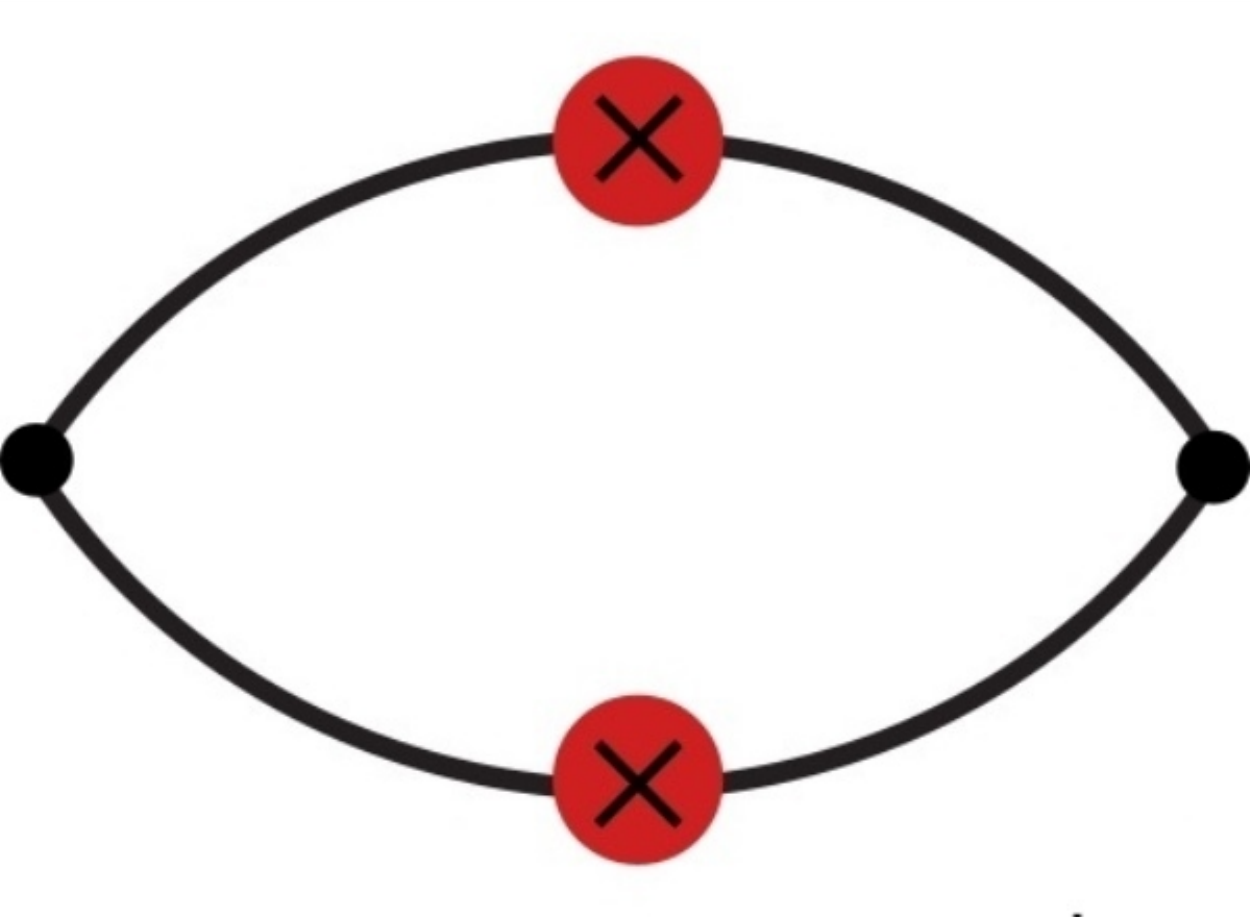}}}
\newcommand{\moned}{\raisebox{-0.4\totalheight}{\includegraphics[scale=0.135]{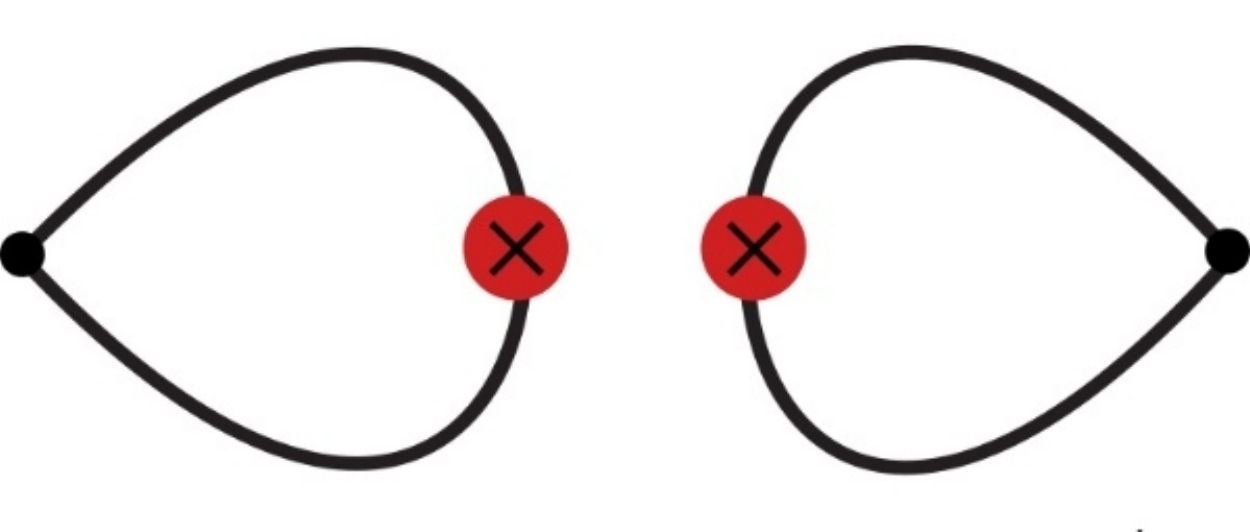}}}

\begin{document}

\title{Lattice determination of the QCD low-energy constant \texorpdfstring{$\ell_{\scriptscriptstyle{7}}$}{l7}}

\author{Claudio Bonanno}
\email{claudio.bonanno@csic.es}
\affiliation{Instituto de F\'isica Te\'orica UAM-CSIC, c/ Nicol\'as Cabrera 13-15, Universidad Aut\'onoma de Madrid, Cantoblanco, E-28049 Madrid, Spain}

\author{Gilberto Colangelo}
\email{gilberto@itp.unibe.ch}
\affiliation{Albert Einstein Center for Fundamental Physics, Institute for Theoretical Physics, University of Bern, Sidlerstrasse 5, 3012 Bern, Switzerland}

\author{Francesco D'Angelo}
\email{francesco.dangelo@roma3.infn.it}
\affiliation{INFN, Sezione di Roma Tre, Via della Vasca Navale 84, I-00146 Rome, Italy}

\author{Massimo D'Elia}
\email{massimo.delia@unipi.it}
\affiliation{Dipartimento di Fisica, Università di Pisa \& INFN, Sezione di Pisa, Largo Pontecorvo 3, I-56127 Pisa, Italy}

\author{Roberto Dionisio}
\email{roberto.dionisio@phd.unipi.it}
\affiliation{Dipartimento di Fisica, Università di Pisa \& INFN, Sezione di Pisa, Largo Pontecorvo 3, I-56127 Pisa, Italy}

\author{Roberto Frezzotti}
\email{roberto.frezzotti@roma2.infn.it}
\affiliation{Dipartimento di Fisica, Università di Roma ``Tor Vergata'' \& INFN, Sezione di Roma Tor Vergata, Via della Ricerca Scientifica 1, I-00133 Rome, Italy}

\author{Giuseppe Gagliardi}
\email{giuseppe.gagliardi@roma3.infn.it}
\affiliation{Dipartimento di Matematica e Fisica, Università ``Roma Tre'' \& INFN, Sezione di Roma Tre, Via della Vasca Navale 84, I-00146 Rome, Italy}

\author{Vittorio Lubicz}
\email{vittorio.lubicz@uniroma3.it}
\affiliation{Dipartimento di Matematica e Fisica, Università ``Roma Tre'' \& INFN, Sezione di Roma Tre, Via della Vasca Navale 84, I-00146 Rome, Italy}

\author{Guido Martinelli}
\email{guido.martinelli@roma1.infn.it}
\affiliation{Dipartimento di Fisica, Università di Roma ``La Sapienza'' \& INFN, Sezione di Roma La Sapienza, Piazzale Aldo Moro 5, I-00185 Rome, Italy}

\author{Francesco Sanfilippo}
\email{francesco.sanfilippo@roma3.infn.it}
\affiliation{INFN, Sezione di Roma Tre, Via della Vasca Navale 84, I-00146 Rome, Italy}

\author{Silvano Simula}
\email{simula@roma3.infn.it}
\affiliation{INFN, Sezione di Roma Tre, Via della Vasca Navale 84, I-00146 Rome, Italy}

\date{\today}

\begin{abstract}
We provide a non-perturbative determination of the scheme- and scale-independent low-energy constant $\ell_{\scriptscriptstyle{7}}$, appearing in the QCD effective chiral Lagrangian at next-to-leading order, by means of lattice QCD simulations with $N_{\scriptscriptstyle{\rm f}}=2+1$ quark flavors. We adopt staggered fermions and extract $\ell_{\scriptscriptstyle{7}}$ from the pion mass splitting by suitably generalizing the method introduced in [\textit{Phys.~Rev.~D} \textbf{104} (2021) 074513] for the Wilson discretization. Adopting 12 gauge ensembles with 3 different values of the pion mass, and 4 different values of the lattice spacing, we are able to achieve controlled extrapolations towards the continuum, infinite volume, and chiral limits. Our final result $\ell_{\scriptscriptstyle{7}} \,\times \, 10^3 = 2.79(58)_{\scriptscriptstyle{\rm stat}}(19)_{\scriptscriptstyle{\rm syst}} = 2.79(61)_{\scriptscriptstyle{\rm tot}}$ agrees with and substantially improves on previous determinations.
\end{abstract}

\maketitle

\section{Introduction}

Effective field theories represent a powerful approach for studying Quantum Field Theories in regimes where a full description across all energy scales is either impractical or unnecessary. They indeed provide a framework which allows one to capture the relevant dynamics at the desired energy scales through approximations that are, in principle, systematically improvable.~In practice, effective theories are built as an expansion in one or more small quantities, and since the number of free parameters typically grows rapidly with the order of the expansion, this can potentially limit their applicability. However, there are many phenomenologically-relevant cases where already the first few terms allow, in certain regimes, a satisfying approximation of very complex theories with a controllable number of parameters, the non-perturbative low-energy domain of Quantum Chromo-Dynamics (QCD) being a paramount example.

In the limit of $\Nf$ massless quarks, the QCD Lagrangian exhibits a global $\mathrm{SU}(\Nf)_{\scriptscriptstyle{\rm L}} \times \mathrm{SU}(\Nf)_{\scriptscriptstyle{\rm R}}$ flavor symmetry, which is spontaneously broken to $\mathrm{SU}(\Nf)_{\scriptscriptstyle{\rm V}}$ at energy scales much smaller than $\Lambda_{\QCDsub}$~\cite{Goldstone:1962es}. This fact allows a low-energy description of strong interactions by means of Chiral Perturbation Theory ($\ChPT$), an effective theory describing massive QCD as a small perturbation around the chirally-symmetric point with massless quarks.

The $\ChPT$ Lagrangian is constructed by writing down all possible terms consistent with the symmetries of the theory in terms of the Nambu--Goldstone boson fields associated to chiral symmetry breaking, ordered in powers of the squared momenta $p^2$ and of the quark masses $m_{\rm q}$ (the so-called $p$-expansion). Each order of this expansion introduces a set of parameters known as Low Energy Constants (LECs), which encode the non perturbative dynamics of the full theory, and must be fixed either phenomenologically from experimental data or through first-principles lattice QCD calculations. At Leading Order (LO) in $p^2$ and $m_{\rm q}$, the $\ChPT$ Lagrangian contains only two LECs: the pion decay constant $F$ and the quark condensate $\Sigma$. Instead, at Next-to-Leading Order (NLO) $\mathcal{O}(p^4)$, and for $\Nf=2$ light quark flavors, one needs to introduce 7 more LECs, $\{\ell_{\scriptscriptstyle{i}}\}_{i\,=\,1,...,7}$. The value of all LECs depends on the strange quark mass $\mqs$.

Among them, the scheme and scale independent LEC $\ls$ is the only one parametrizing strong isospin-breaking (IB) effects at NLO. For instance, it enters the strong charged/neutral pion mass splitting~\cite{Gasser:1983yg,Gasser:1984gg}:
\beq\label{eq:mass_split}
M^2_{\pi^{+}}-M^2_{\pi^{0}} = (\Delta m)^2 \frac{8B^2}{F^2}\ls\left[1 + \mathcal{O}(\mql)\right],
\eeq
where ($\mqu$, $\mqd$ are the up, down quark masses)
\beq\label{eq:def_deltam}
\mql \equiv \frac{1}{2}(\mqd + \mqu), \qquad \Delta m \equiv \frac{1}{2}(\mqd - \mqu),
\eeq
$B\equiv\Sigma/F^2$, and where we are adopting the convention $F_\pi^{\physsup} \simeq 92$ MeV for physical quark masses. Thus, $\ls$ plays a crucial role in several physical processes where IB effects are important, ranging from hadron physics to Beyond Standard Model phenomenology. Despite its importance, among the seven $\Nf=2$ NLO $\ChPT$ LECs, $\ls$ is currently affected by the largest uncertainties, and the number of papers tackling its calculation is rather scarce.\\
The first phenomenological estimate of $\ls$ was given in \cite{Gasser:1983yg, Gasser:1984gg} using arguments related to the $\eta-\pi_0$ mixing:
\beq
\ls = \frac{F_\pi^2}{6M_\eta^2}\sim 5 \times 10^{-3}.
\eeq
However, NLO corrections turn out to be quite large and of the same order of LO contributions, thus giving rise to sizable systematics~\cite{GrillidiCortona:2015jxo}: $\ls = 7(4) \times 10^{-3}$.\\
On the other hand, also the current status of first-principles determination of $\ls$ from numerical Monte Carlo simulations of lattice QCD is not very advanced (for a review on the status of the lattice determinations of the QCD LECs see Sec.~5 of Ref.~\cite{FlavourLatticeAveragingGroupFLAG:2021npn}). The RBC-UKQCD Collaboration provided  a {first \emph{indirect}} determination {of $\ls$} in Ref.~\cite{Boyle:2015exm} through a global fit involving several pseudoscalar masses and decay constants: $\ls = 6.5(3.8) \times 10^{-3}$. The accuracy of this determination is however comparable with the phenomenological estimate discussed above.

Improving the accuracy of current determinations of $\ls$ has been recognized as a necessary ingredient for accurate predictions about axion physics. The axion mass $m_a$ receives a NLO correction whose overall uncertainty is at the percent level, with the most significant contribution coming from the one on $\ls$~\cite{GrillidiCortona:2015jxo,Gorghetto:2018ocs}. Moreover, the dominant source of error in the determination of the axion quartic self-coupling $\lambda_a$ (about 6\%) comes from $\ls$ too~\cite{GrillidiCortona:2015jxo}. Finally, recent studies of axion-pion scattering within $\mathrm{SU}(2)$ $\ChPT$ at NLO have shown that the error on $\ls$ induces a 15 -- 20\% uncertainty in the $a\pi \to \pi\pi$ amplitude, impacting constraints on the axion mass obtained from hot Dark Matter bounds~\cite{DiLuzio:2021vjd,DiLuzio:2022gsc,DiLuzio:2022tbb}.

Going beyond the present state of the art demands devising dedicated non-perturbative lattice strategies to effectively determine $\ls$. Recently, progress in this direction has been achieved. More precisely, two \emph{direct} lattice methods to compute $\ls$ from lattice QCD have been proposed in Ref.~\cite{Frezzotti:2021ahg}, both rooted on the RM123 approach~\cite{deDivitiis:2011eh,Giusti:2017dmp}. One is based on the calculation of the coupling between the neutral pion and the iso-symmetric pseudoscalar quark density; the other, which is the one we will adopt in this study, is instead based on the calculation of the pion mass splitting in Eq.~\eqref{eq:mass_split}. The main idea behind this strategy is to evaluate the charged and neutral pion mass difference $M^2_{\pi^+} - M^2_{\pi^0}$ at order $\mathcal{O}[(\Delta m)^2]$ by expanding the path-integral around the isosymmetric point $\mqu=\mqd\equiv\mql$ in powers of the quarks mass difference $\Delta m$. This makes it straightforward to directly evaluate derivatives of the pion mass difference with respect to $\Delta m$, which are indeed proportional to $\ls$ itself, as we will discuss later. In the same paper~\cite{Frezzotti:2021ahg}, the authors also provided a proof-of-principle lattice calculation using rotated Twisted Mass fermions~\cite{Frezzotti:2021slr} for a single gauge ensemble, and were able to reduce the bound on $\ls$ down to $\ls = 2.5(1.4)\times10^{-3}$ combining the results from the two methods.

Given the very promising results obtained in~\cite{Frezzotti:2021ahg}, in this work we aim at performing a dedicated lattice study targeting the determination of $\ls$ using the same pion mass splitting method of~\cite{Frezzotti:2021ahg}, but implementing it with staggered fermions. This choice is motivated by the computational efficiency of the staggered discretization, which is significantly less expensive to simulate compared to other fermion formulations; moreover, checking the consistency 
of determinations obtained adopting different fermion discretizations represents, as usual, a solid test of universality and of the claimed control over the systematic effects which may affect lattice QCD results.\\
Clearly, staggered fermions also present additional complications related to the breaking of taste symmetry at finite lattice spacing which require particular attention. Despite these challenges, we were able to generalize the strategy of~\cite{Frezzotti:2021ahg} to staggered quarks, and apply it to several staggered gauge ensembles obtaining reliable results for $\ls$. This allowed us to obtain controlled continuum and chiral extrapolations of this quantity for the first time, and to attain an overall final uncertainty which significantly improves on previous estimates.

This paper is organized as follows: in Sec.~\ref{sec:setup} we describe our numerical setup and the mass-splitting method we adopted to compute $\ls$; in Sec.~\ref{sec:res} we present our results for $\ls$; finally, in Sec.~\ref{sec:conclu} we draw our conclusions.

\section{Numerical setup}\label{sec:setup}

\subsection{Lattice QCD discretization and simulation parameters}\label{discretization}

\begin{table*}[!t]
\centering
\begin{tabular}{|c|c|c|c|c|c|c|c|c|c|c|}
\hline
$R \equiv \mql/\mqs^{\physsup}$ & $\beta$ & $\Ns$ & $a \mql$ & $a \mqs$ & $a$ [fm]& $N_{\rm conf}$ & $M_\pi$ [MeV] & $F_\pi$ [MeV] & $L=a\Ns$ [fm] & $M_\pi L$ \\
\hline
\hline
\multirow{4}{*}{$0.1421$}
& 3.67838 & 20 & 0.008828 & 0.062136 & 0.1515 & 6200 & 260(3) &98.5(2.1) & 3.03 & 4.00 \\
& 3.75000 & 24 & 0.007150 & 0.050300 & 0.1265 & 5400 & 262(3) &99.0(2.3) & 3.04 & 4.08 \\
& 3.86847 & 32 & 0.005383 & 0.037884 & 0.0964 & 3373 & 263(4) &98.9(2.2) & 3.08 & 4.21 \\
& 3.98775 & 40 & 0.004244 & 0.029867 & 0.0758 & 3160 & 269(5) &98.4(2.3) & 3.03 & 4.13 \\
\hline
\hline
\multirow{4}{*}{$0.2131$}
& 3.67838 & 20 & 0.013242 & 0.062136 & 0.1532 & 1800 & 320(2) & 102.4(2.7) & 3.06 & 4.96 \\
& 3.75000 & 24 & 0.010722 & 0.050300 & 0.1278 & 1800 & 315(3) & 101.8(2.5) & 3.07 & 4.90 \\
& 3.86847 & 32 & 0.008075 & 0.037884 & 0.0976 & 2730 & 315(5) & 100.9(2.4) & 3.12 & 4.98 \\
& 3.98775 & 40 & 0.006366 & 0.029867 & 0.0764 & 1390 & 317(6) & 101.1(2.5) & 3.06 & 4.92 \\
\hline
\hline
\multirow{4}{*}{$0.3197$}
& 3.67838 & 20 & 0.019863 & 0.062136 & 0.1556 & 1320 & 381(1) & 106.8(2.6) & 3.11 & 6.00 \\
& 3.75000 & 24 & 0.016080 & 0.050300 & 0.1297 & 1280 & 382(2) & 105.8(2.4) & 3.11 & 6.02 \\
& 3.86847 & 32 & 0.012112 & 0.037884 & 0.0989 & 1250 & 380(1) & 105.5(2.5) & 3.16 & 6.09 \\
& 3.98775 & 40 & 0.009549 & 0.029867 & 0.0768 & 1320 & 386(4) & 106.3(2.6) & 3.07 & 6.01 \\
\hline
\end{tabular}   
\caption{Summary of the simulation parameters of the gauge ensembles employed in this study. The LCP parameters, including pion masses and lattice spacings, were obtained in~\cite{Bonanno:2023xkg} starting from the physical point LCP determined in~\cite{Aoki:2009sc, Borsanyi:2010cj, Borsanyi:2013bia}. The values of the lattice spacing have been determined in~\cite{Bonanno:2023xkg} with the $w_0$ scale setting approach~\cite{BMW:2012hcm} with a $\sim 2\%$ error at most, and their value in fm units was obtained assuming $w_0 = 0.1757(12)~\mathrm{fm}$~\cite{BMW:2012hcm} for all ensembles, independently of the pion mass. The variable $N_{\rm conf}$ is the number of thermalized gauge configurations generated for each ensemble, each separated by 10 RHMC updating steps. For each ensemble we also report the values of $M_\pi$ and $F_\pi$.}
\label{tab:lattice_params}
\end{table*}

The numerical setup (i.e., the lattice QCD discretization and the algorithm employed for the Monte Carlo sampling of the path integral) and the gauge ensembles adopted here are both inherited from the previous study~\cite{Bonanno:2023xkg}.

We discretize QCD on an hypercubic $\Ns^4$ space-time lattice, with periodic (antiperiodic) boundary conditions for gluons (quarks). We adopt a tree-level Symanzik-improved action for the gluonic sector:
\beq
\begin{aligned}
S_{\YMsub}^{\Lsup}[U]=-\frac{\beta}{3}\sum_{x,\mu\neq\nu}
&\left\{\frac{5}{6} \Re\mathrm{Tr}\left[\Pi_{\mu\nu}^{(1\times1)}(x)\right]+\right.\\
&\left.-\frac{1}{12} \Re\mathrm{Tr}\left[\Pi_{\mu\nu}^{(1\times2)}(x)\right]\right\}, 
\end{aligned}
\eeq
with $\Pi_{\mu\nu}^{(n\times m)}(x)$ being the product of gauge links along $n\times m$ rectangular paths starting at point $x$ and extending in the $(\mu,\nu)$ plane. The fermionic sector consists of $\Nf=2+1$ flavors of rooted stout-smeared staggered fermions. The fermion matrix is given by:
\beq
\mathcal{M}_{\rm f}^{\stagsup}[U]\equiv D_{\stagsub}[U^{(2)}]+ a\mf,
\eeq
\beq
\begin{aligned}
D_{\stagsub}[U^{(2)}]=\sum_{\mu\,=\,1}^{4}\eta_{\mu}  (x)&\left[U_{\mu}^{(2)}(x) \delta_{x,y-\hat{\mu}} \right.\\
&\left.-U_{\mu}^{(2)}{}^{\dagger}(x-\hat{\mu}) \delta_{x,y+\hat{\mu}}\right],
\end{aligned}
\eeq
\beq
\eta_{\mu}(x)=(-1)^{x_{1}+\dots+x_{\mu-1}},
\eeq
with $\eta_\mu(x)$ the so called staggered phases, $a$ the lattice spacing, $\mf$ the bare mass of the quark flavor f, and where the stouted links $U^{(2)}_{\mu}(x)$ are taken after 2 steps of isotropic stout smearing with $\rho_{\rm{stout}}=0.15$~\cite{Morningstar:2003gk}.\\
The total partition function:
\beq
\begin{aligned}
Z_{\LQCDsub}=\int[\mathrm{d}U] \, \mathrm{e}^{-S_{\YMsub}^{\Lsup}[U]} \times &\det\left\{\mathcal{M}_{\ell}^{\stagsup}[U]\right\}^{\frac{1}{2}}\\
\times&\det\left\{\mathcal{M}_{\rm s}^{\stagsup}[U]\right\}^{\frac{1}{4}},
\end{aligned}
\eeq
with $\ell$ and s labeling the light and strange quark flavors respectively, has been sampled with the Rational Hybrid Monte Carlo (RHMC) algorithm~\cite{Clark:2006wp, Clark:2006fx}.

In this work we will employ 12 different gauge ensembles, whose simulation parameters are reported in Tab.~\ref{tab:lattice_params}. These ensembles correspond to 3 different Lines of Constant Physics (LCPs) where the lattice size is fixed to $ L \equiv a\Ns \simeq 3~\mathrm{fm}$, the bare strange quark mass $\mqs$ is tuned to keep its renormalized value constant and equal to its physical value $\mqs^{\physsup}$, and the bare mass of the degenerate light doublet $\mqu=\mqd=\mql$ is varied to explore different values of the pion mass $M_\pi$. These LCPs were determined in Ref.~\cite{Bonanno:2023xkg}, starting from the results of Refs.~\cite{Aoki:2009sc, Borsanyi:2010cj, Borsanyi:2013bia} for the bare parameters belonging to the physical point LCP, by varying the bare light quark mass at fixed bare strange quark mass. More precisely, we considered light-to-strange quark mass ratios $R\equiv \mql/\mqs^{\physsup}$ equal to 4,~6 and 9 times the physical ratio $R^{\physsup} = \mql^{\physsup}/\mqs^{\physsup}$ determined in Refs.~\cite{Aoki:2009sc, Borsanyi:2010cj, Borsanyi:2013bia}. These choices correspond to $M_\pi^2 \simeq 4,~6,~9 \, {M_{\pi}^{\physsup}}^2$, $M_\pi^{\physsup} = 135$ MeV, and $M_\pi L \ge 4$, a value that is sufficient to keep finite-size effects below our statistical errors, as we shall discuss later. For each LCP we considered 4 values of the lattice spacing $a$ ranging from $\sim 0.15$ fm to $\sim 0.075$ fm. Our setup thus allows us to take continuum limits $a\to 0$ at fixed light and strange quark masses, and then to take the SU(2) chiral limit $\mql\to0$ at fixed physical strange quark mass $\mqs^{\physsup}$, i.e., $R\to 0$. Further details about the determination of the LCP simulation parameters can be found in the original paper~\cite{Bonanno:2023xkg}.\footnote{The results for $F_\pi$ in Tab.~\ref{tab:lattice_params} correct those originally given in Ref.~\cite{Bonanno:2023xkg}, whose calculation was affected by a mismatch in the renormalization constants, resulting in a $\sim 15-19\%$ shift upward. This issue regarded the analysis of the ensembles at heavier-than physical pions, and it does not affect the chiral limit of $F_\pi$ given in~\cite{Bonanno:2023xkg}, but only its slope as a function of $M_\pi$. The latter was not discussed in~\cite{Bonanno:2023xkg}, and it is reported for the first time in the present work in Appendix~\ref{app:l3_l4}.}

\subsection{\texorpdfstring{$\ls$}{l7} from the mass-splitting method}\label{sec:methods}

The mass-splitting method of~\cite{Frezzotti:2021ahg} relies on two main ingredients: the fact that the strong charged/neutral pion mass difference induced by strong IB effects is parametrized by $\ls$ via Eq.~\eqref{eq:mass_split}; and the fact that this formula can be directly evaluated at the first non-trivial order in $\Delta m$ by using the RM123 method.

The starting point is to rewrite~\eqref{eq:mass_split} in a way that is more suitable for a numerical evaluation. By expanding the left-hand side at LO in $\Delta m$ and using the $\ChPT$ LO relation $M^2 = B (\mqu + \mqd) = 2 B \mql$, we have:
\beq
\begin{aligned}
\label{eq:first_step_l7_NLO}
M^2_{\pi^{+}} - M^2_{\pi^{0}} &\simeq 2 M \left( M_{\pi^{+}} - M_{\pi^{0}} \right)\\
&= (\Delta m)^{2} \frac{2M^4}{\mql^2 F^2} \ls,
\end{aligned}
\\[0.5em]
\label{eq:l7_MM_def}
\implies \ls = \frac{\mql^{2}F^{2}}{M^{3}} \times \frac{(M_{\pi^{+}}-M_{\pi^{0}})_{\QCDsub}}{(\Delta m)^{2}},
\eeq
where $(M_{\pi^{+}}-M_{\pi^{0}})_{\QCDsub}$ denotes the mass splitting arising solely from the up/down quark mass difference, i.e., excluding $\mathcal{O}(\alpha_{\rm em})$ electromagnetic effects. Thus, at the order in $\Delta m$ at which we are working, the computation of $\ls$ is reduced to the computation of:
\beq\label{eq:mass_split_derivative}
\hspace*{-1.4\baselineskip}
\frac{(M_{\pi^{+}}-M_{\pi^{0}})_{\QCDsub}}{(\Delta m)^{2}} = \frac{1}{2}(M''_{\pi^{+}}-M''_{\pi^{0}})_{\isoQCDsub} + \mathcal{O}(\Delta m)
\eeq
in isosymmetric QCD (i.e., the theory with $\Delta m = 0$), with $\prime$ denoting differentiation with respect to $\Delta m$.

Such quantity can be extracted on the lattice from the second derivative with respect to $\Delta m$ of the difference $C_{\pi^+\pi^+}(t)-C_{\pi^0\pi^0}(t)$ between the Euclidean time correlation functions of the charged and neutral pions. As a matter of fact, the asymptotic behavior of these correlators for large Euclidean time separations is:
\begin{equation}\label{eq:corrs}
\begin{aligned}
C_{\pi^{+} \pi^{+}}(t) & \underset{{a\ll t\ll T}}{\sim} A_{\pi^{+}} \cosh \left[M_{\pi^{+}}\left(\frac{T}{2}-t\right)\right], \\
C_{\pi^0 \pi^0}(t) & \underset{{a\ll t\ll T}}{\sim} A_{\pi^0} \cosh \left[M_{\pi^0}\left(\frac{T}{2}-t\right)\right],
\end{aligned}
\end{equation}
where $T=a\Ns$ is the Euclidean time extent of the lattice, and $A_{\rm P}$ represents the amplitude for the particle P obtained from the interpolating operator $\mathcal{O}_{\rm P}$,
\beq
A_{\rm P} &=& \frac{\vert Z_{\mathrm{P}\mathrm{P}} \vert^2}{M_{\rm P}}\mathrm{e}^{-M_{\rm P} \frac{T}{2}},\\
Z_{\mathrm{P}\mathrm{P}} &=& \langle \mathrm{P} \vert \mathcal{O}_{\mathrm{P}} \vert 0 \rangle.
\eeq
Therefore, by taking derivatives with respect to $\Delta m$ one obtains~\cite{Frezzotti:2021ahg}:
\beq\label{eq:diff_cors_asymptotic}
\begin{aligned}
&\frac{\left[C_{\pi^{+} \pi^{+}}^{\prime \prime}(t)-C_{\pi^0 \pi^0}^{\prime \prime}(t)\right]_{\isoQCDsub}}{C_{\pi \pi}^{\isoQCDsup}(t)} \underset{{a\ll t\ll T}}{\sim}\\ & \frac{A_{\pi^{+}}^{\prime \prime}-A_{\pi^0}^{\prime \prime}}{A_\pi}+\left(M_{\pi^{+}}^{\prime \prime}-M_{\pi^0}^{\prime \prime}\right) \left(\frac{T}{2}-t\right)\\
&\times \tanh \left[M_\pi\left(\frac{T}{2}-t\right)\right]
\end{aligned}
\eeq
with $C_{\pi \pi}^{\isoQCDsup}(t)$ the usual pion correlator computed in isosymmetric QCD.

To obtain the numerator of the left-hand side of Eq.~\eqref{eq:diff_cors_asymptotic} from an actual lattice QCD Monte Carlo calculation and extract $\left(M_{\pi^{+}}^{\prime \prime}-M_{\pi^0}^{\prime \prime}\right)$, we follow the RM123 approach~\cite{deDivitiis:2011eh}, which allows one to evaluate derivatives with respect to $\Delta m$ by expanding around isoQCD. Let us briefly introduce this method by starting from the fermionic part of the Euclidean QCD Lagrangian for a non-degenerate light quark doublet $\psi = (\psi_{\usub},\psi_{\dsub})$:
\beq\label{eq:quark_lagrangian}
\mathcal{L}_{\rm q}=\overline{\psi} \left( \gamma_\mu D_\mu  + \mathcal{M}\right)\psi,
\eeq
where $\mathcal{M}$ is the mass matrix, diagonal in flavor space, i.e., $\mathcal{M} = \mathrm{diag}(\mqu,\mqd)$. The mass term can be written as a sum of a $\mathrm{SU}(2)$ symmetric part and a term which explicitly breaks the isospin symmetry:
\beq
\begin{aligned}
\mathcal{L}_{m}&=\frac{\mqu+\mqd}{2}\left(\overline{\psi}_{\usub}\psi_{\usub}+\overline{\psi}_{\dsub}\psi_{\dsub}\right)\\
& \quad -\frac{\mqd-\mqu} {2}\left(\overline{\psi}_{\usub}\psi_{\usub}-\overline{\psi}_{\dsub}\psi_{\dsub}\right)\\
&=\mql\left(\overline{\psi}_{\usub}\psi_{\usub}+\overline{\psi}_{\dsub}\psi_{\dsub}\right)-\Delta     m\left(\overline{\psi}_{\usub}\psi_{\usub}-\overline{\psi}_{\dsub}\psi_{\dsub}\right)\\
&=\mql \, \overline{\psi}\psi-\Delta m\,\overline{\psi}\tau_{3}\psi,
\end{aligned}
\eeq
with $\tau_3=\mathrm{diag}(1,-1)$ being the third Pauli matrix. This allows us to isolate the IB contribution $\mathcal{L}_{\IBsub}$ within the starting Lagrangian~\eqref{eq:quark_lagrangian} as:
\beq
\mathcal{L}_{\rm q} = \underbrace{\overline{\psi} \left(\gamma_\mu D_\mu + \mql\right)\psi}_{\mathcal L_0}  -\underbrace{\Delta m \, \overline{\psi} \tau_3\psi}_{\mathcal L_{\IBsub}}.
\eeq
Within the RM123 approach, $\mathcal{L}_{\IBsub}$ is treated as a perturbation with respect to $\mathcal{L}_0$, and the path integral is expanded around the iso-symmetric point $\mqu=\mqd\equiv\mql$. The vacuum expectation value (VEV) of a generic observable $\mathcal{O}$ can then be computed as follows:
\beq
\begin{aligned}
\braket{\mathcal{O}} = \braket{\mathcal{O}}_{0} &- \underbrace{\braket{\mathcal{O}S_{\IBsub}}_0}_{\mathcal{O}(\Delta m)} \\
& + \underbrace{\frac{1}{2} \left[\braket{\mathcal{O}S^2_{\IBsub}}_0 - \braket{\mathcal{O}}_0 \braket{S^2_{\IBsub}}_0 \right]}_{\mathcal{O}[(\Delta m)^2]} \\ &+ \mathcal{O}[(\Delta m)^3],\\
\end{aligned}
\eeq
where $\langle \dots\rangle_0$ represents the VEV with respect to the isosymmetric theory (recall that $\braket{S_{\IBsub}}_0 = \braket{\int \mathrm{d}^4 x \, \mathcal{L}_{\IBsub}}_0 = 0$).

Using the RM123 approach outlined so far, one can expand the difference between the charged/neutral pion correlators in powers of $\Delta m$. Performing the relevant Wick contractions, one obtains the following diagrammatic expansion at the first non-trivial order~\cite{Frezzotti:2021ahg}:
\beq\label{eq:pion_mass_diff_diagrams}
\begin{aligned}
&C_{\pi^{+}\pi^{+}}(t) - C_{\pi^{0}\pi^{0}}(t)=\\[10pt]
&= 2\left({Z_{\mathrm{S}}Z_m}\right)^{2}(\Delta m)^{2} \\
& \quad\, \times \left[\,\,\monec\,\, - \,\, \moned \,\,\right]\\[12pt]
&\equiv 2 \left( {Z_{\mathrm{S}}Z_m}\right)^{2}(\Delta m)^{2}\left[ C_{\fourptsub}^{\connsup}(t) - C_{\fourptsub}^{\discsup}(t)\right]\\
&= 2(\Delta m)^2 \left[ C_{\fourptsub}^{\connsup}(t) -      C_{\fourptsub}^{\discsup}(t)\right],
\end{aligned}
\eeq
where we have denoted with $C_{\fourptsub}^{\connsup}(t)$ and $C_{\fourptsub}^{\discsup}(t)$ respectively the connected and disconnected 4-point correlation functions arising at $\mathcal{O}[(\Delta m)^2]$ in the RM123 expansion of $C_{\pi^{+}\pi^{+}}(t) - C_{\pi^{0}\pi^{0}}(t)$.\\
In the diagrammatic representation of these 4-point correlators, the full dots in the previous diagrams denote an insertion of $\gamma_5$, while the crossed ones correspond to an insertion of $S_{\IBsub}=\int \mathrm{d}^4 x \, \mathcal{L}_{\IBsub}$. We also included the renormalization constants for the mass and for the scalar density $Z_m,Z_{\mathrm{S}}$ due, respectively, to the factors of $\Delta m$ and $\overline{\psi}\tau_3\psi$ in $\mathcal{L}_{\IBsub}$. However, within our staggered fermion discretization, $Z_m=Z_{\mathrm{S}}^{-1}$, thus the quantity in~\eqref{eq:pion_mass_diff_diagrams} is a renormalization group invariant~\cite{Bali:2013coa}.

Given that, at the first non-trivial order in $\Delta m$,
\beq
\begin{aligned}
C_{\pi^{+} \pi^{+}}(t)-C_{\pi^0 \pi^0}(t) &= \frac{(\Delta m)^2}{2}\\
&\times \left[C_{\pi^{+} \pi^{+}}''(t)-C_{\pi^0 \pi^0}''(t)\right]_{\Delta m\,=\,0},
\end{aligned}
\eeq
we recognize that:
\beq
\begin{aligned}
\frac{1}{2} \left[C_{\pi^{+} \pi^{+}}''(t)-C_{\pi^0 \pi^0}''(t)\right]_{\Delta m\,=\,0} = \\
2\left[C_{\fourptsub}^{\connsup}(t) - C_{\fourptsub}^{\discsup}(t)\right].
\end{aligned}
\eeq
In the end, we can construct the following lattice estimator for $\ls$:
\beq\label{eq:l7_estimator}
\begin{aligned}
\ls^{\effsup}(t) &\equiv \frac{2F_\pi^2 m^2_{\lsub}}{M_\pi^3} \frac{1}{\mathcal{F}\left(\frac{T}{2}-t,M_\pi\right)}\left[\frac{\delta C}{C}(t)-\frac{\delta C}{C}(t-a)\,\right],\\
\\[0.01em]
\frac{\delta C}{C}(t) &\equiv \frac{\left[C_{\fourptsub}^{\connsup}(t)-C_{\fourptsub}^{\discsup}(t)\right]_{\isoQCDsub}}{C_{\pi\pi}^{\isoQCDsup}(t)},\\
\\[0.01em]
\mathcal{F}(x,M) &\equiv x\tanh(Mx) - (x+a)\tanh\left[M(x+a)\right],
\end{aligned}
\eeq
with $a$ the lattice spacing, and $M_\pi$, $F_\pi$ the finite-quark-mass determinations of these quantities in Tab.~\ref{tab:lattice_params}. The function $\ls^{\effsup}(t)$, for asymptotically large time separations, will tend to $\ls$.

Finally, on the technical side, both the 2-point and the 4-point correlators entering the estimator~\eqref{eq:l7_estimator} were computed using standard stochastic methods, using temporal dilution with one stochastic source per time slice in order to invert the staggered Dirac operator. We also verified in a few cases that increasing the number of sources gave compatible error estimates.

\section{Numerical Results}\label{sec:res}

\subsection{Lattice determination of \texorpdfstring{$\ls$}{l7}}\label{sec:calc_l7}
In this section we will exemplify the lattice determination of $\ls$ from the mass-splitting method introduced in Sec.~\ref{sec:methods} for one gauge ensemble.

Let us start by discussing the evaluation of the correlation functions entering the $\ls$ estimator in Eq.~\eqref{eq:l7_estimator}. As it is well known, see, e.g., Refs.~\cite{Golterman:1984dn,Ishizuka:1993mt,Borsanyi:2010cj}, when using staggered fermions there are several choices for the pion interpolating operator,
\beq
\mathcal{O}^{\Ssup}_\pi = \overline{\psi} \gamma_5 \otimes \xi_{\Ssub} \psi,
\eeq
each one corresponding to a different spin-taste structure: S = (I, P, V, A, T), labeling respectively the scalar, pseudoscalar, vector, axial-vector and tensor channels. Due to explicit taste-breaking at finite lattice spacing, the ground state masses in each of these channels differ among themselves, but they all converge to the same value in the continuum limit (see Appendix~\ref{app:pion_masses} for more details on this point). However, at finite lattice spacing, only the pseudoscalar channel, $\xi_{\Psub} = \xi_{5}$, corresponds to an actual pseudo-Nambu--Goldstone boson whose mass is protected by the remnant of the continuum chiral symmetry enjoyed by staggered fermions. Thus, this is customarily the preferred choice for the pion interpolating operator in most lattice calculations involving the staggered discretization. As a matter of fact, for the isoQCD 2-point pion correlation function and the extraction of $M_\pi$ and $F_\pi$, whose calculation was tackled in the previous study~\cite{Bonanno:2023xkg}, and to which we refer for more details, this was indeed the case.

\begin{figure*}[!t]
\centering
\includegraphics[scale=0.25]{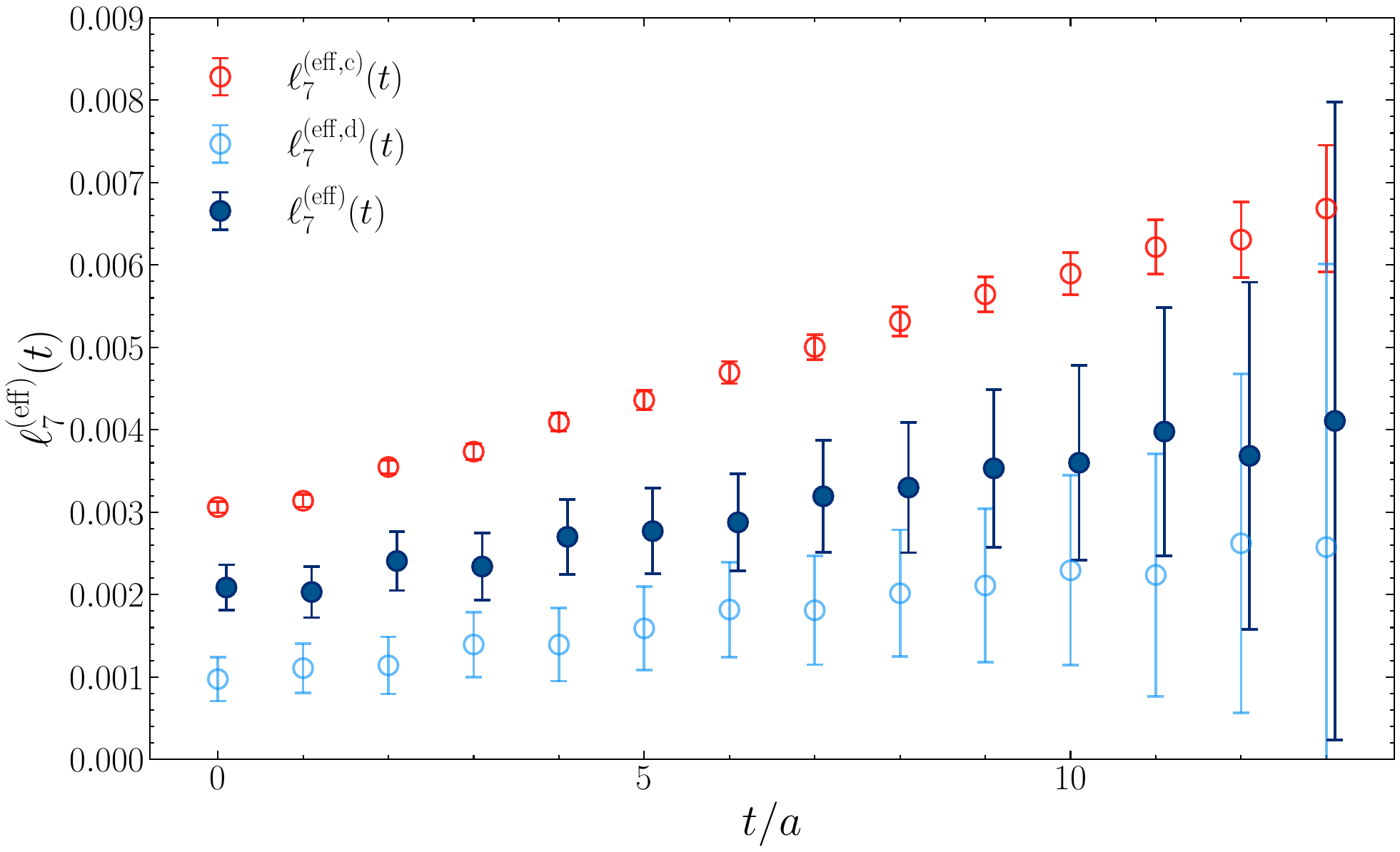}
\includegraphics[scale=0.25]{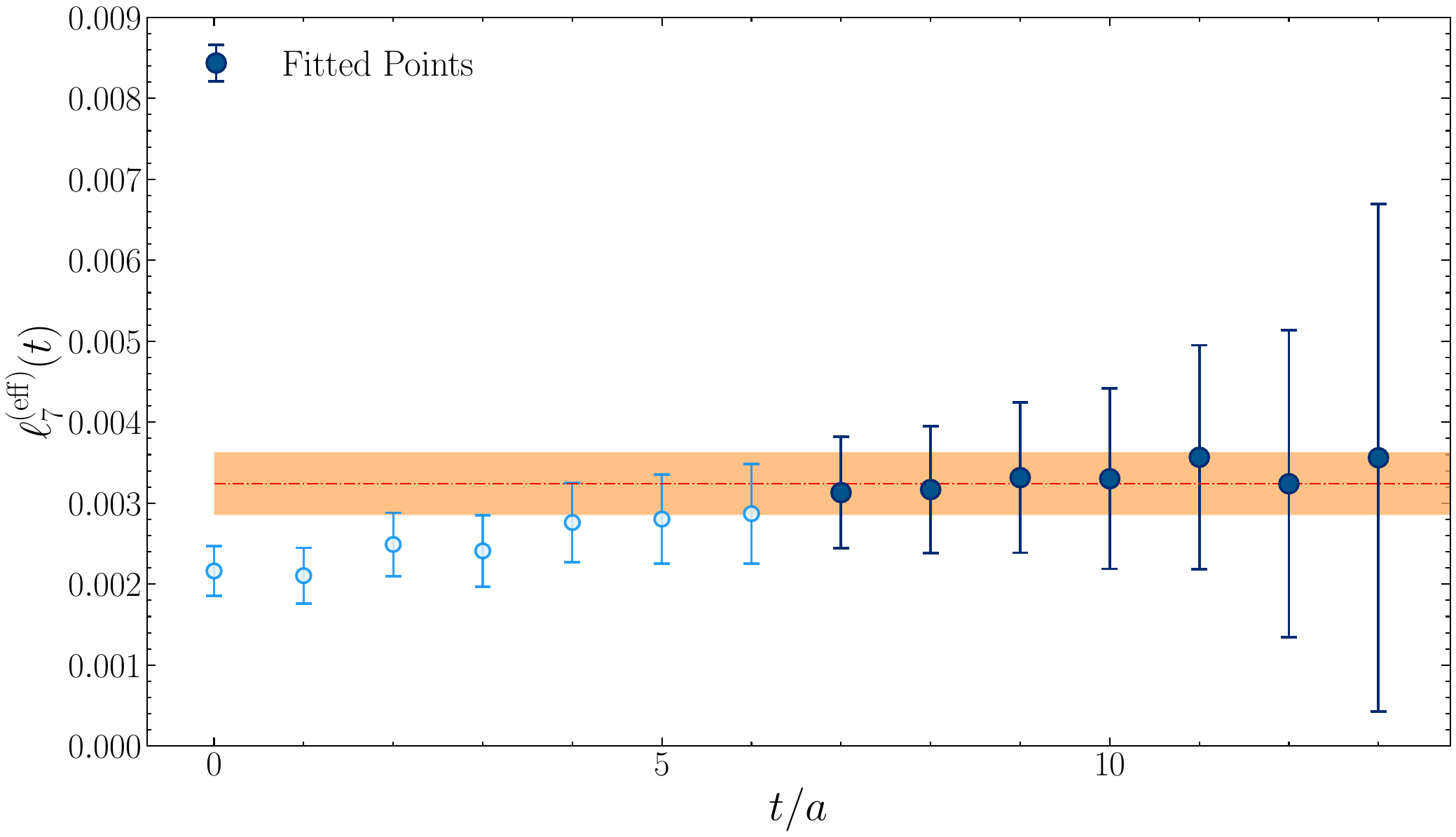}
\caption{Left panel: lattice estimator $\ls^{\effsup}(t)$ and individual contributions of the connected/disconnected diagrams to this function for the ensemble with $R\simeq0.1421$ and $a=0.0964$ fm. Right panel: extraction of $\ls$ from the large-time-separation plateau exhibited by $\ls^{\effsup}(t)$.}
\label{fig:plateau}
\end{figure*}

Instead, concerning the 4-point correlators $C_{\fourptsub}^{\connsup}$ and $C_{\fourptsub}^{\discsup}$, we found that the standard pseudoscalar channel revealed to be impractical, as with this choice of the pion interpolating operator the disconnected correlator turned out to be completely dominated by statistical noise, and remained compatible with zero within errors for all time separations\footnote{A similar behavior for disconnected diagrams in the pseudoscalar channel was observed in Ref.~\cite{Aubin:2007mc}. In that study it is pointed out that a term appearing in the effective Staggered $\ChPT$ Lagrangian allows disconnected contributions in the meson propagator only in the singlet, vector, and axial vector channels. Thus, to study the $\eta$ and $\eta^\prime$ mesons, where disconnected diagrams are crucial due to their flavor-singlet nature, the $\gamma_5 \times \mathrm{I}$ interpolating operator was employed, as it is the only pseudoscalar bilinear with a non-vanishing disconnected contribution in the staggered case.}. Thus, the pseudoscalar channel does not allow for a reliable extraction of $\ls$, motivating the need to explore alternative spin-taste structures. The first natural candidate beyond the pseudoscalar channel is the axial-vector one, corresponding to $\xi_{\Asub}=\xi_{5\mu}$. In this case, both the connected and disconnected diagrams exhibit non-zero signal with reasonable statistical quality, thus allowing a reliable calculation of the estimator $\ls^{\effsup}(t)$ (see Appendix~\ref{app:comp_channels} for a comparison of the obtained results for the disconnected contribution to $\ls$ from the pseudo-scalar and the axial channels). As a further check, we also verified that compatible results for $\ls$ are obtained from the vector channel ($\xi_{\Vsub} = \xi_{\mu}$) once the continuum limit is taken, although within larger error bars, see Appendix~\ref{app:comp_channels}, thus motivating our choice of the axial-vector channel.

In Fig.~\ref{fig:plateau} (left panel), we show the time-dependence of the estimator $\ls^{\effsup}(t)$ for the ensemble with $\mql = 4\mql^{\physsup}$ and lattice spacing $a \simeq 0.0964$ fm. In order to illustrate how the physical signal for $\ls$ arises from the difference of the connected and disconnected correlation functions of Eq.~\eqref{eq:pion_mass_diff_diagrams}, we also plotted in the same figure the individual contributions to $\ls^{\effsup}(t)$ of the connected (c) and disconnected (d) diagrams, which are such that:
\beq
\ls^{\effsup}(t) = \ls^{\effcsup}(t) - \ls^{\effdsup}(t).
\eeq
As it can be appreciated from Fig.~\ref{fig:plateau} (right panel), the estimator exhibits a plateau for large time separations, as expected.

\noindent More precisely, for the case at hand, we observe the onset of the plateau for $t\ge8a = T/4$. It is reassuring to find the onset of the plateau for $F_\pi$ and $M_\pi$ in similar time ranges, cf.~for example Fig.~6 in Ref.~\cite{Bonanno:2023xkg}. The evaluation of $\ls$ is thus performed via a constant best fit in a region close to $T/2$, where contamination of excited states is minimized, and the ground-state dominance is expected. The time window for the best fit was chosen by ensuring stability of the central value upon variation of the lower/upper range, and reasonable reduced $\chi^2$, while the final error was assessed via a binned bootstrap analysis; typical windows are $[t_{\min},t_{\max}]$ with $t_{\min}/T\sim 0.2-0.3$ and $t_{\max}/T\sim 0.4-0.45$. The obtained values of $\ls$ for all ensembles are collected in Tab.~\ref{tab:res}.

We conclude by reporting for comparison the result from the mass splitting method obtained in Ref.~\cite{Frezzotti:2021ahg} with twisted mass Wilson fermions, for a lattice spacing $a \simeq 0.095$ fm, a pion mass $M_\pi \simeq 260$ MeV, and a volume $M_\pi L \simeq 4$:
\beq
\ls \times 10^3 = 3.5(2.0), \qquad \text{(Ref.~\cite{Frezzotti:2021ahg})}.
\eeq
Among our ensembles we have one with the same values of the lattice spacing, of the pion mass and of the lattice volume, namely, our next-to-finest lattice spacing ensemble for the smallest value of $R$ simulated, cf.~Tab.~\ref{tab:lattice_params}. With staggered fermions in this case we find:
\beq
\ls \times 10^3 = 3.25(38), \qquad \text{(this study)}.
\eeq
It should be noted that the error reduction achieved here cannot be entirely ascribed to the different sample sizes, as our statistics is larger by about a factor of 3 with respect to the one of~\cite{Frezzotti:2021ahg}, but we find a smaller uncertainty by about a factor of 5. \footnote{On the other hand, the statistical accuracies of $F_\pi$ and $M_\pi$ obtained with staggered quarks here and with twisted mass Wilson fermions in~\cite{Frezzotti:2021ahg} are similar. More comprehensive investigations with other discretizations and possibly with mixed action setups are needed to clarify how the statistical accuracies of different types of observables depend on the chosen fermion formulation.}

\subsection{Finite-size effects on \texorpdfstring{$\ls$}{l7}}\label{sec:FSE}

Concerning finite-size effects, as outlined in Sec.~\ref{sec:setup}, we chose for all our simulations a lattice size of $L=a\Ns\simeq3$ fm, corresponding to $M_\pi L \simeq 4,5,6$ for the 3 sets of ensembles with $M_\pi\simeq 265,316,384$ MeV we considered.

\begin{figure}[!t]
\centering
\includegraphics[scale=0.44]{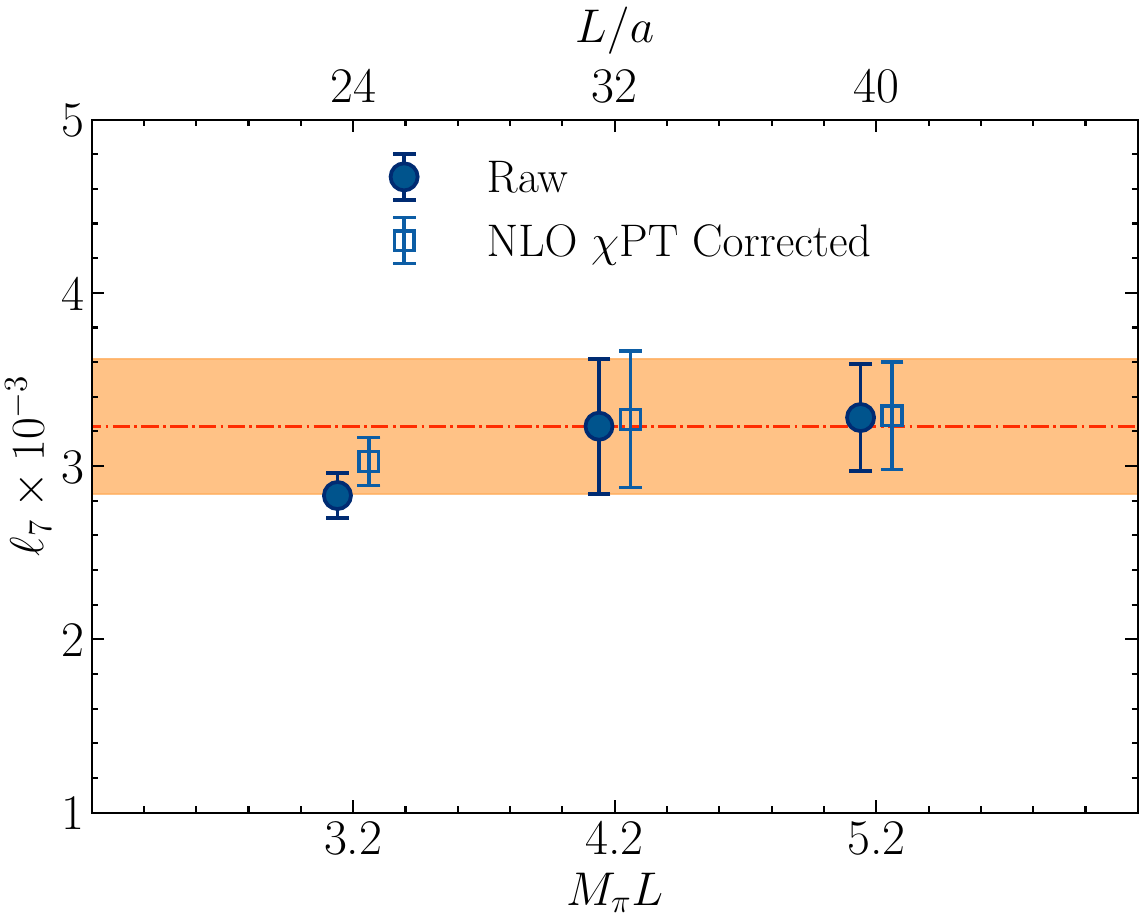}
\caption{Study of finite-size effects on $\ls$ for our ensemble with $a=0.0964$ $\mathrm{fm}$ and $M_\pi=263(4)$ $\mathrm{MeV}$. They are invisible within our statistical errors for $M_\pi L \ge 4$. The same conclusions are reached after subtracting finite-size effects estimated from NLO $\ChPT$: $\ls(L)\to\ls(L)/[1+R_{\ls}(L)]$, see Sec.~\ref{sec:FSE}.}
\label{fig:FSE}
\end{figure}

For the next-to-finest lattice spacing of our lightest ensembles with $M_\pi L\simeq 4$ we explicitly checked that this threshold is sufficient to keep finite-size effects under control by performing two additional simulations for a smaller and a larger value of $M_\pi L\simeq3$ and $5$. The result of this test, shown in Fig.~\ref{fig:FSE}, indeed confirm that our choice for the lattice volume is safe. Such conclusion is also supported by analytical results obtained from NLO $\ChPT$. Indeed, analytic expressions for the volume-dependence of $M_\pi$ and $F_\pi$ are available up to NLO in the chiral expansion~\cite{Colangelo:2004xr,Colangelo:2005gd} (for the pion mass, also NNLO effects are known~\cite{Colangelo:2006mp} and have been included in our calculation):
\beq
M_\pi(L) &=& M_\pi\left[1+R_{M_\pi}(L)\right],\\
F_\pi(L) &=& F_\pi\left[1+R_{F_\pi}(L)\right],\\
\implies \ls(L) &=& \ls  \left\{1+2\left[R_{F_\pi}(L)-R_{M_\pi}(L)\right]\right\},
\eeq
with $R_{M_\pi}$ and $R_{F_\pi}$ given by Eqs.~(26) and (27) of~\cite{Colangelo:2005gd}.

We find that $R_{\ls}(L) \equiv 2\left[R_{F_\pi}(L)-R_{M_\pi}(L)\right]$ is about $R_{\ls}(L)\simeq -6.5\%,-1.2\%,-0.3\%$ for $M_\pi L \simeq 3,4,5$, cf.~Fig.~\ref{fig:FSE}. Based on the difference between NLO and NNLO results for $R_{M_\pi}$, we estimate that the reported numbers for $R_{\ls}$ are affected by at most a $\sim 1\%$ error due to the truncation of the chiral expansion. Since finite-volume effects estimated from $R_{\ls}(L)$ are at least one order of magnitude smaller than our statistical errors on $\ls$ for $M_\pi L \ge 4$, we conclude that they are completely negligible with our current precision.

\begin{table}[!t]
\centering
\begin{tabular}{|c|c|c|c|}
\hline
&&&\\[-1em]
$R \equiv \mql/\mqs^{\physsup}$ & $a$ [fm] & $\ls \times 10^{3}$ & $w_0^2\Delta$\\
\hline
\hline
\multirow{4}{*}{0.1421} & 0.1515 & 1.11(37) & 0.0587 \\
& 0.1265 & 2.18(25) & 0.0400 \\
& 0.0964 & 3.25(38) & 0.0203 \\
& 0.0758 & 3.67(40) & 0.00990 \\
\hline
\hline
\multirow{4}{*}{0.2131} & 0.1532 & 3.18(89) & 0.0611 \\
& 0.1278 & 3.56(48) & 0.0409 \\
& 0.0976 & 4.36(43) & 0.0208 \\
& 0.0764 & 4.62(38) & 0.00994 \\
\hline
\hline
\multirow{4}{*}{0.3197} & 0.1556 & 4.73(88) & 0.0648 \\
& 0.1297 & 5.16(57) & 0.0433 \\
& 0.0989 & 6.21(51) & 0.0215 \\
& 0.0768 & 6.08(36) & 0.0101 \\
\hline
\end{tabular}
\caption{Summary of all direct lattice determinations of $\ls$ as a function of $a$ and of the parameter $R$ (proportional to the light quark masses $\mql$) using the mass-splitting method. We also report the squared pion mass splitting $\Delta\equiv M_\pi^2(\xi_{\Asub})-M_\pi^2(\xi_{\Psub})$ between the axial-vector and the pseudo-scalar channel in units of $1/w_0^2$, which will be used for the study of systematic errors, see Sec.~\ref{sec:cont_lim_l7}. The error on $w_0^2\Delta$ is of the order of $\sim 2\%$ at most.}
\label{tab:res}
\end{table}

\subsection{Continuum extrapolations of \texorpdfstring{$\ls$}{l7}}\label{sec:cont_lim_l7}

\begin{figure*}[!t]
\centering
\includegraphics[scale=0.266]{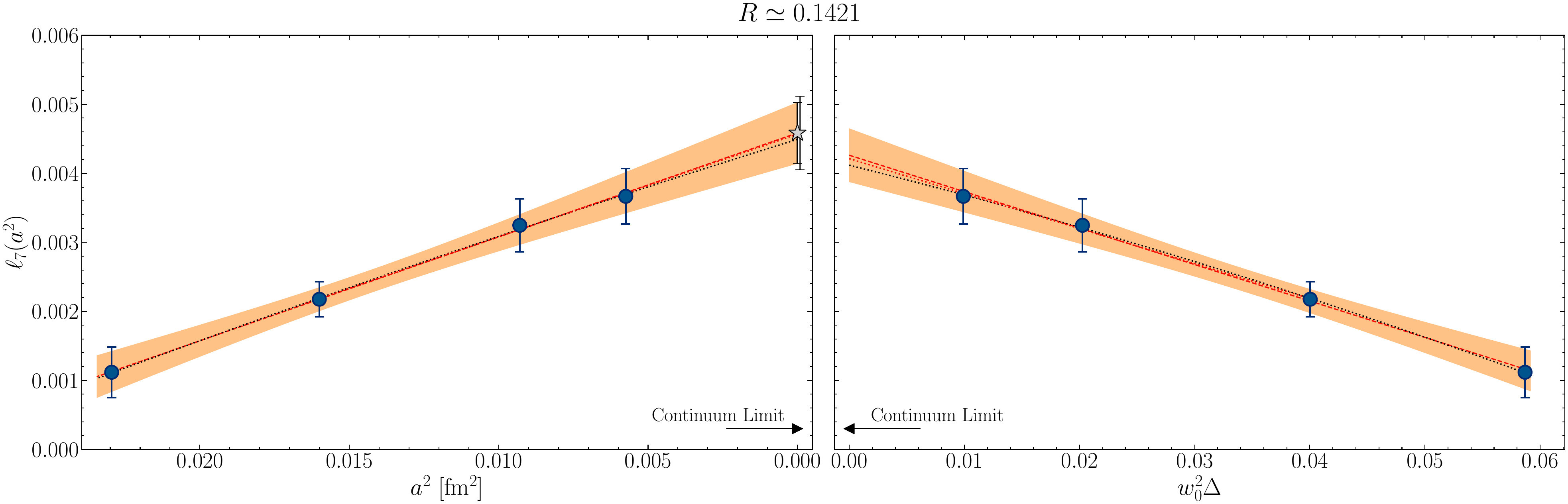}
\includegraphics[scale=0.266]{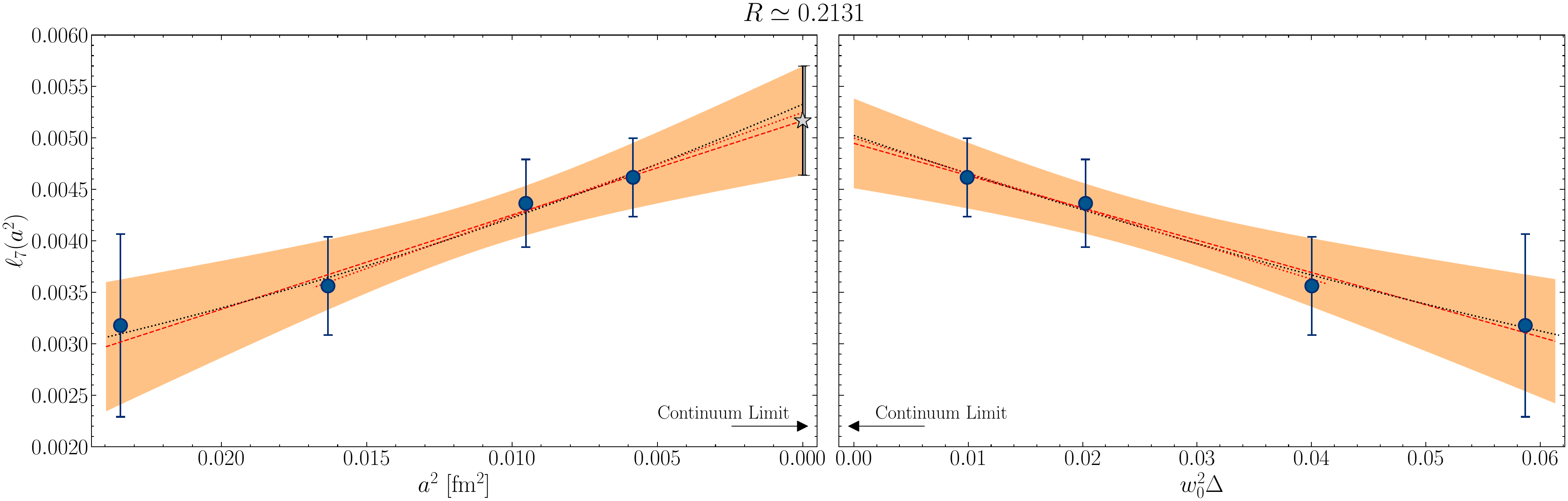}
\includegraphics[scale=0.266]{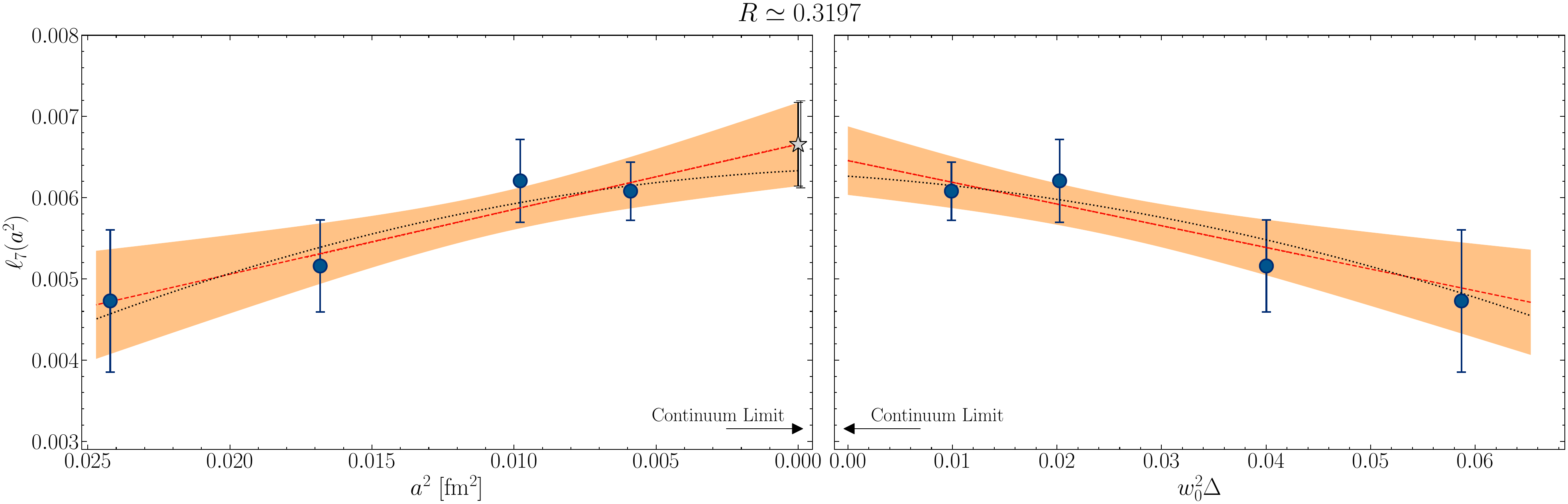}
\caption{Continuum limit extrapolations of $\ls$ for the 3 values of the parameter $R=\mql/\mqs^{\physsup}$ explored. Left panels report best fits as a function of $a^2$, while right panels report best fits as a function of $w_0^2\Delta$. Continuous lines (with their related continuous shaded areas) correspond to linear fits using all available data points, dashed lines represent linear fits restricted to the 3 finest lattice spacings, and dotted lines represent quadratic fits. The star points represent our final results for the continuum limits, and are obtained from the linear extrapolations in $a^2$ (no star point is represented for extrapolations in $w_0^2\Delta$, as these fits are only used to assess systematic errors). Since the total error bar, given by the sum in quadrature of the statistical and systematic uncertainties, is barely distinguishable from the statistical one for the two largest values of $R$, it has been slightly shifted horizontally for clarity.}
\label{fig:continuum}
\end{figure*}

After the extraction of $\ls$ at finite lattice spacing and quark mass, we have all the necessary ingredients to perform continuum and chiral extrapolations.

Let us start from the continuum limit, which is taken in all cases at fixed value of the light and strange quark mass, i.e., at fixed value of $R=\mql/\mqs^{\physsup}$. The continuum extrapolations were carried out by fitting our data with a few Ans\"atze:
\begin{align*}
\ls(a^2, R) &= \ls(R) + C_2(R) a^2, &\quad (\mathrm{A}1),\\
\ls(a^2, R) &= \ls(R) + C_2(R) a^2 + C_4(R) a^4,&\quad (\mathrm{A}2),\\
\ls(\Delta, R) &= \ls(R) + C_2^\prime(R) w_0^2 \Delta,&\quad (\mathrm{A}3),\\
\ls(\Delta, R) &= \ls(R) + C_2^\prime(R) w_0^2\Delta + C_4^\prime(R) w_0^4 \Delta^2,&\quad (\mathrm{A}4),
\end{align*}
where for the fit Ans\"atze A1 and A3 we performed 2 fits excluding/including the coarsest lattice spacing. In all formulas, $\ls(R)$ denotes the continuum-extrapolated value, while $\Delta$ denotes the taste-splitting between the squared axial-vector and pseudo-scalar pion masses:
\beq
\Delta \equiv M_\pi^2(\xi_{\Asub})-M_\pi^2(\xi_{\Psub}),
\eeq
whose values in units of $1/w_0$ are reported in Tab.~\ref{tab:res}. The fit Ans\"atze A3 and A4 are rooted on Symanzik effective theory (SYMEFT), which predicts that leading-order $a^2$ lattice artifacts are modified as $a^2 \alpha^n_{\rm s}(\mu=1/a)$~\cite{Husung:2019ytz,Husung:2025nsv}, with $\alpha_{\rm s}(\mu)$ the running strong coupling, and $n$ depending on the particular sea and valence discretizations considered. Our choice of the quantity $w_0^2\Delta$ follows the proposal of Ref.~\cite{Boccaletti:2024guq} [cf.~Eqs.~(17) and (18) in Sec.~4 of that study], where the authors study hadronic contributions to the anomalous muon magnetic moment with stout-smeared rooted staggered fermions. In particular, the authors observe that $\Delta$ vanishes approaching the continuum limit with a rate which is approximately $\Delta(a) \underset{a\,\to\,0}{\sim} a^2\alpha^3_{\rm s}(\mu=1/a)$, and thus use this quantity in their SYMEFT-inspired fit Ans\"atze to study systematic errors related to the continuum limit.

As it can be appreciated from Fig.~\ref{fig:continuum}, our data can be nicely fitted with $\mathcal{O}(a^2)$ corrections both when including/excluding the data at the coarsest lattice spacing, all giving perfectly compatible results within errors. Similar conclusions are drawn including a further $a^4$ term, or adopting the fit Ans\"atze based on SYMEFT.\\
Given the overall good agreement among all possible continuum limit fit functions, in all cases we considered the outcome of the linear fit in $a^2$, performed considering all lattice spacings, to assign a central value and a statistical error to our continuum determinations of $\ls(R)$.
In addition to statistical uncertainties, we also assigned to our continuum extrapolations a systematic error related to the small difference observed when extrapolating our data with different fit Ans\"atze. Following the procedure outlined in \cite{ExtendedTwistedMassCollaborationETMC:2022sta,Bonanno:2023xkg}, we computed:
\beq
\Delta_{\mathrm{A}}=\frac{\left|\left[\ls(R)\right]_{{\mathrm{A}1}}-\left[\ls(R)\right]_{{\mathrm{A}}}\right|}{\sqrt{\Delta^2_{{\mathrm{stat}}}\left[\ls(R)\right]_{{\mathrm{A}1}}+\Delta^2_{{\mathrm{stat}}}\left[\ls(R)\right]_{{\mathrm{A}}}}},
\eeq
which quantifies the relative difference between the central values obtained with the fit Ansatz A1 (including all lattice spacings) and another fit Ansatz ``A'' among the ones earlier listed.\\

\noindent The systematic error was then estimated as:
\beq\label{eq:syst_error}
\Delta_{\mathrm{syst,A}}=\left|\left[\ls(R)\right]_{\mathrm{A}1}-\left[\ls(R)\right]_{\mathrm{A}}\right|\mathrm{erf}\left(\frac{\Delta_{\mathrm{A}}}{\sqrt{2}}\right),
\eeq
where $\mathrm{erf}(x)$ denotes the standard error function,
\beq
\mathrm{erf}(x) = \frac{2}{\sqrt{\pi}} \int_0^x \mathrm{d}t \, \mathrm{e}^{-t^2}.
\eeq

\noindent The quantity~\eqref{eq:syst_error} represents the mismatch between the central values of the two fit Ans\"atze, weighted with an estimation of the probability that this mismatch is not due to a statistical fluctuation. All sources of systematic errors are then summed in quadrature: $\Delta_{\rm syst} = \sqrt{\sum_{\rm A} \Delta^2_{\rm A}}$. The final continuum extrapolations for each value of $R$, including systematic errors, are collected in Tab.~\ref{tab:final}. As it can be appreciated, systematic errors are always sub-dominant with respect to statistical errors: even for $R\simeq 0.1421$, which has the largest systematic error, the statistical error is still $\sim 85\%$ of the total budget (computed as the quadrature sum of the two uncertainties).

\subsection{Chiral extrapolation of \texorpdfstring{$\ls$}{l7}}\label{sec:chir_lim_l7}

\begin{table*}[!t]
\centering
\begin{tabular}{|c|c|c|c|c|c|c|}
\hline
&&&&&&\\[-1em]
$R\equiv \mql / \mqs^{\physsup}$ & $\ls(R)\times 10^{3}$ & $M_\pi(R)$ [MeV] & $F_\pi(R)$ [MeV] & $c_1(R)$ & $c_2(R)$ & $\tilde{\ell}_{\scriptscriptstyle{7}}(R) \times 10^3$\\
&&&&&&\\[-1em]
\hline
0.3197 & 6.66(51)$_{\mathrm{stat}}$(15)$_{\mathrm{syst}}$ & 384(7) & 105.1(2.9) & 0.637(61) & 0.5541(42) & {2.37(29)$_{\mathrm{stat}}$(04)$_{\mathrm{syst}}$}\\
0.2131 & 5.17(53)$_{\mathrm{stat}}$(07)$_{\mathrm{syst}}$ & 316(6) & 100.7(2.8) & 0.762(52) & 0.6125(37) & {2.42(27)$_{\mathrm{stat}}$(03)$_{\mathrm{syst}}$}\\
0.1421 & 4.58(45)$_{\mathrm{stat}}$(29)$_{\mathrm{syst}}$ & 265(6) & 97.7(3.1)& 0.857(61) & 0.6715(34) & {2.64(32)$_{\mathrm{stat}}$(12)$_{\mathrm{syst}}$}\\
\hline
\end{tabular}
\caption{Continuum limits of $\ls$ as a function of the light-to-strange quark mass ratio $R$. We also report the continuum limits of $M_\pi(R)$ and $F_\pi(R)$ (reported errors include both statistical and systematic uncertainties), and of the improved estimator $\tilde{\ell}_{\scriptscriptstyle{7}}(R)$ where leading logs have been removed via $c_1(R)$ and $c_2(R)$ (see Sec.~\ref{sec:cont_lim_l7} for more details).}
\label{tab:final}
\end{table*}

We are now ready to extrapolate our continuum results at finite quark mass towards the SU(2) chiral limit, $\mql\to0$ at fixed $\mqs=\mqs^{\physsup}$, which in our setup corresponds to $R = \mql / \mqs^{\physsup}\to0$.

To achieve a solid control on the chiral extrapolation, it is fundamental to know the analytic form of the leading finite-quark mass contributions to $\ls$. This in turn requires the knowledge of the Next-to-Next-to-Leading-Order (NNLO) contributions to the pion mass splitting in $\chi$PT. Despite the proliferation of operators entering the effective Lagrangian at this order of the chiral expansion, there are only a few isospin-breaking operators at $\mathcal{O}(p^6)$~\cite{Bijnens:1999hw}. This makes it feasible to work out the structure of NNLO contributions to Eq.~\eqref{eq:mass_split}, and thus the finite quark mass corrections to Eq.~\eqref{eq:l7_MM_def}.

The calculation of the strong pion mass splitting at NNLO in $\ChPT$ is fully carried out for the first time in this study. In the following, we will report the results that are necessary for the chiral extrapolation of $\ls$, while further details can be found in Appendix~\ref{app:NNLO_pion_mass_splitting}.

Remarkably, all NNLO contributions to the pion mass splitting are $\mathcal{O}\left[(\Delta m)^2\right]$, and their structure turns out to be rather simple, as there are only two $\mathcal{O}(\mql)$ terms appearing in Eq.~\eqref{eq:pion_mass_splitting_NNLO}. One is a chiral logarithm of order $\mathcal{O}(\mql \log \mql)$, which turns out to be proportional to $\ls$ itself---see also~\cite{Gasser:1980sb,Gasser:1982ap}, where the calculation of the leading non-analytic contributions to $\Delta M_\pi^2$, once matched to our notation, yields a chiral log multiplying $\ls$ with the same coefficient as ours. The second is a linear $\mathcal{O}(\mql)$ correction, whose coefficient depends instead on some combination of the couplings of the NNLO isospin-breaking operators appearing at $\mathcal{O}(p^6)$, see Appendix~\ref{app:NNLO_pion_mass_splitting}. All other corrections are higher-order in the chiral expansion, i.e., $\mathcal{O}(\mql^2)$ in the quark mass. In the end, Eqs.~\eqref{eq:mass_split} and~\eqref{eq:l7_MM_def} are modified respectively as follows:
\begin{widetext}
\begin{equation}\label{eq:pion_mass_splitting_NNLO}
M^2_{\pi^+}-M^2_{\pi^0} = (\Delta m)^2 \frac{8 B^2}{F^2} \times \left\{ \ls \left[1 -3\frac{M^2}{16\pi^2 F^2} \log\left(\frac{M^2}{16\pi^2 F^2}\right)\right] + C M^2 + \mathcal{O}\left(M^4\right) \right\},
\end{equation}
\begin{equation}\label{eq:l7_def_NNLO}
\frac{M_\pi}{M}\frac{\mql^{2}F^{2}}{M^{3}} \frac{M_{\pi^{+}}-M_{\pi^{0}}}{(\Delta m)^{2}} = \ls \left[1 - 3\frac{M^2}{16\pi^2 F^2} \log\left(\frac{M^2}{16\pi^2 F^2}\right)\right] + C M^2 + \mathcal{O}\left(M^4\right),
\end{equation}
\end{widetext}

\noindent Note that, to pass from \eqref{eq:pion_mass_splitting_NNLO} to \eqref{eq:l7_def_NNLO}, we here used $\Delta M_\pi^2 \simeq 2 M_\pi \Delta M_\pi$, as opposed to $\Delta M_\pi^2 \simeq 2 M \Delta M_\pi$ in Eq.~\eqref{eq:first_step_l7_NLO}, since we are now retaining $\mathcal{O}(\mql)$ corrections to $\ls$.

\begin{figure*}[!t]
\centering
\includegraphics[scale=0.4]{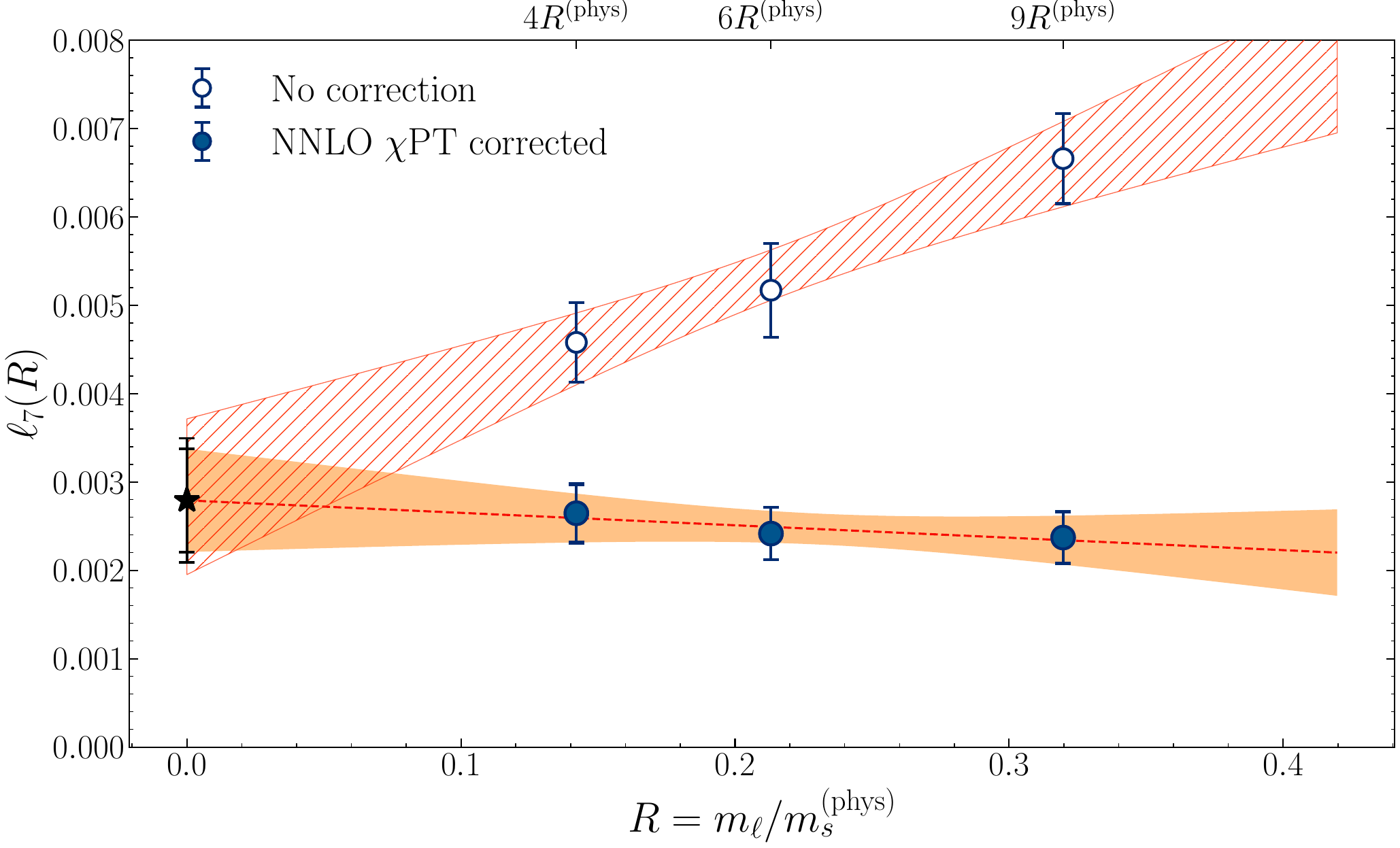} 
\caption{Chiral limit extrapolation of the corrected values of $\ls(R)$ according to a linear function of the ratio $R=\mql/\mqs^{\physsup}$. The full star point at $R=0$ stands for the extrapolated value in the chiral limit, according to the best fit of the corrected data (dashed line). The two error bars plotted for the chiral extrapolation represent, respectively, the statistical and the sum of the statistical and systematic uncertainties. For the sake of comparison we also show the non-corrected data and their naive linear extrapolation, depicted as a dashed shaded band.}
\label{fig:chiral_limit}
\end{figure*}

Inspired by Eq.~\eqref{eq:l7_def_NNLO}, before performing any chiral extrapolation, we apply the following correction factors to our continuum determinations of $\ls(R)$ in Tab.~\ref{tab:final},
\beq
\ls(R) \,\,\, \longrightarrow \,\,\, \ls(R) \times c_1(R) \times c_2(R) \equiv \tilde{\ell}_{\scriptscriptstyle{7}}(R),
\eeq
where the two correction factors $c_1(R)$ and $c_2(R)$ read:
\beq
c_1(R) &=& \frac{F^2}{(2B\mqs R)^{2}} \frac{M^4_\pi(R)}{F^2_\pi(R)},\\
c_2^{-1}(R) &=& 1-3\frac{B \mqs R}{8\pi^2F^2}\log \left(\frac{B \mqs R}{8\pi^2F^2}\right).
\eeq
The first factor, $c_1(R)$, takes into account the fact that, for the purpose of evaluating the pre-factor of $\ls^{\effsup}$ in Eq.~\eqref{eq:l7_estimator} we have used the simulated values of $M_\pi^2(R)/\mql$ and $F_\pi(R)$, cf.~Tab.~\ref{tab:lattice_params}, rather than their expression in the chiral limit $M^2/\mql=2B$ and $F$, which is what actually appears in the left-hand side of Eq.~\eqref{eq:mass_split}, see~\cite{Gasser:1982ap,Gasser:1983yg,Gasser:1984gg}. This choice is irrelevant for the purpose of computing the chiral limit of $\ls(R)$, but of course introduces additional finite-quark-mass corrections to Eq.~\eqref{eq:l7_def_NNLO}, that $c_1(R)$ aims to cancel. The second factor $c_2(R)$ aims instead to remove the $\mathcal{O}(\mql\log\mql)$ leading-log term appearing in the right-hand side of Eq.~\eqref{eq:l7_def_NNLO}. To compute $c_1$ and $c_2$, we took $F$ from the latest FLAG average, $F = 86.6(6)$ MeV~\cite{FlavourLatticeAveragingGroupFLAG:2021npn}, and $M^2/R=B\mqs=[2.36(5)\times10^5]~\mathrm{MeV}^2$ from~\cite{Bonanno:2023xkg}. For the finite-quark-mass quantities $M_\pi(R)$ and $F_\pi(R)$, instead, we took the continuum-extrapolated results of~\cite{Bonanno:2023xkg}, reported in Tab.~\ref{tab:final}. The correction factors and the corrected quantity $\tilde{\ell}_{\scriptscriptstyle{7}}$ are also found in Tab.~\ref{tab:final}.

After the application of the two $\chi$PT-inspired corrections outlined above, we expect our corrected data for $\ls(R)$ to be well-described by the following function:
\beq\label{eq:chiral_fit_final}
\tilde{\ell}_{\scriptscriptstyle{7}}(R) = \ls + \tilde{C} R + \mathcal{O}(R^2),
\eeq
i.e., according to a leading linear dependence in $R$, with no $\mathcal{O}(R\log R)$ terms, up to $\mathcal{O}(R^2)$ sub-leading terms. As it can be seen from Fig.~\ref{fig:chiral_limit}, the corrected data show almost a flat behavior as a function of $R$, $\tilde{C}\simeq (-1.4\pm 2.4)\cdot 10^{-3}$, and are thus in perfect agreement with our expectations. A naive linear extrapolation in $R$ of the uncorrected data would still give a  compatible extrapolation, but with a much larger slope.

\begin{figure*}[!t]
\centering
\includegraphics[scale=0.55]{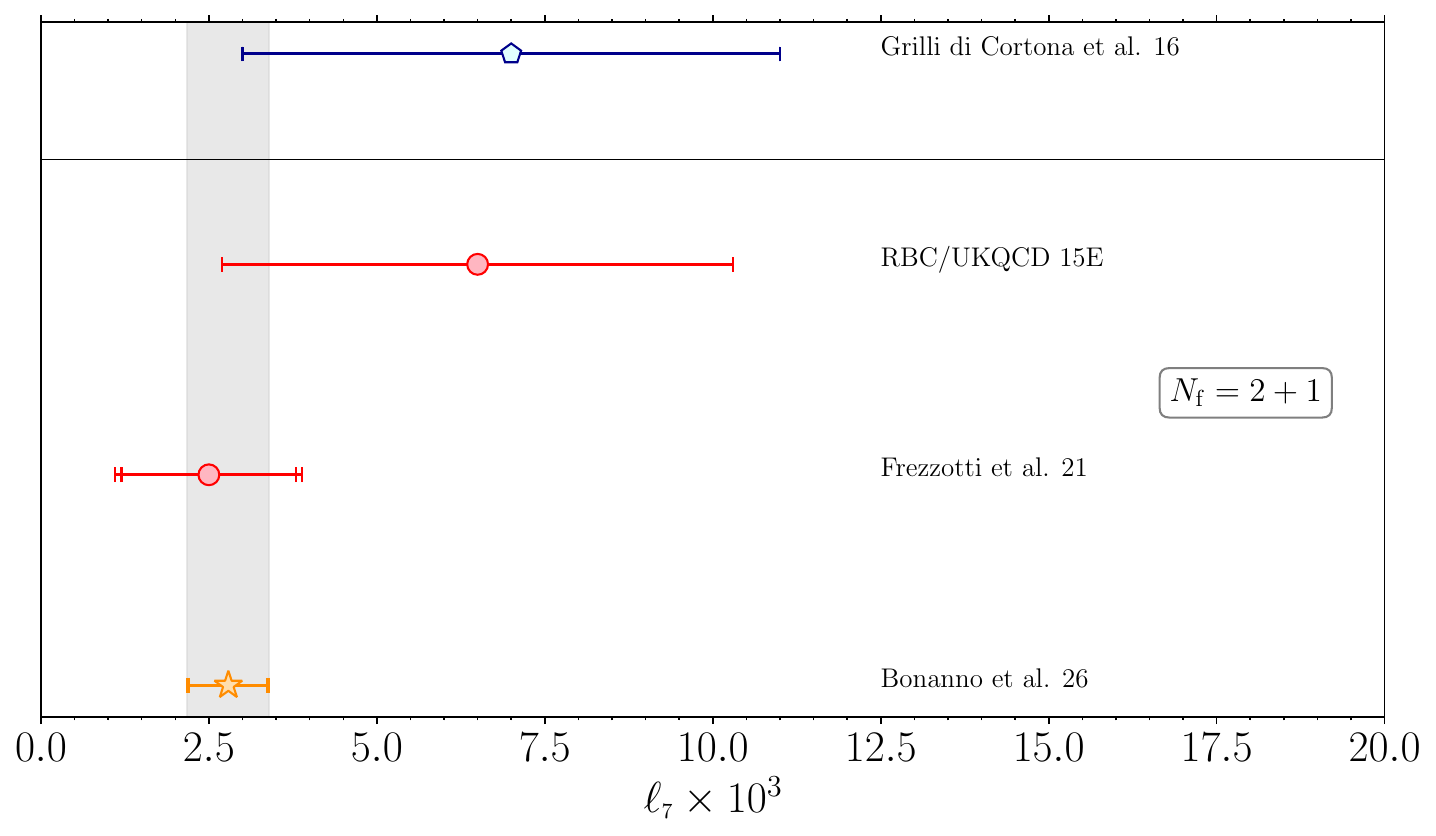}
\caption{Comparison of our $N_{\rm f}=2+1$ lattice determination of $\ls$ (starred point, shaded area) with previous ones from Refs.~\cite{Boyle:2015exm,Frezzotti:2021ahg}, and with the phenomenological estimate of~\cite{GrillidiCortona:2015jxo}.}
\label{fig:comp_l7}
\end{figure*}

In order to propagate systematic uncertainties associated with the continuum limit to the chiral extrapolation, we followed a bootstrap-based strategy. In particular, we performed a bootstrap resampling of each data point, and added to each value of $\ls(R)$ a zero-mean stochastic Gaussian noise with variance equal to the squared systematic error. For each bootstrap sample we repeated the chiral fit, and estimated the final systematic error from the average absolute deviation between the central value of the chiral limits of each bootstrap sample and the central value of the chiral extrapolation obtained ignoring systematic errors. Concerning instead the systematic error related to the possible influence of higher-order $\mathcal{O}(R^2)$ terms, with only three points it is not possible to perform a comprehensive study. A possible crude estimate can however be obtained, using again Eq.~\eqref{eq:syst_error}, by comparing the linear extrapolation presented above with the one obtained from a 0-degrees-of-freedom best fit to the data corresponding to the two smallest values of $R$. This uncertainty is added in quadrature to the one obtained from the propagation of the systematic errors on the continuum limits.

In the end, our final result for the NLO LEC $\ls$ in the chiral limit, including both statistical and systematic uncertainties from continuum and chiral extrapolations, and represented as a full star point in Fig.~\ref{fig:chiral_limit}, reads:
\beq
\ls \times 10^{3} = 2.79(58)_{\mathrm{stat}}(19)_{\mathrm{syst}} = 2.79(61)_{\mathrm{tot}}.
\eeq
As it can be seen, our final error is dominated by statistical uncertainties (mainly due to the chiral extrapolation), which account for about the $85\%$ of the total error budget. Therefore, further reducing it would require determinations of $\ls$ for lower quark masses. Unfortunately, the signal-to-noise ratio of $\ls$ rapidly deteriorates as we approach the physical point, mainly due to the disconnected contribution to $\ls$ (see Appendix~\ref{app:error_scaling} for more details). Thus, determining $\ls$ at, for example, the physical point $R \simeq 0.0355$ would require a very challenging numerical effort.

\section{Conclusions}\label{sec:conclu}

In this work we presented an extensive direct lattice calculation of the QCD low-energy constant $\ls$, which parametrizes IB effects in $\mathrm{SU}(2)$ $\ChPT$ at NLO, adopting staggered fermions.

To this end, we employed a recently-proposed method~\cite{Frezzotti:2021ahg} to extract $\ls$ from the strong charged/neutral pion mass splitting, based on the RM123 approach to compute derivatives with respect to the quark mass difference $\Delta m = (\mqd-\mqu)/2$ from an expansion around the $\Delta m=0$ isosymmetric theory. This method was originally proposed and applied to twisted mass Wilson fermions, and is here implemented for staggered fermions for the first time.

Our calculation utilizes 3 different LCPs with the same value of the strange quark mass (fixed to its physical value), and varying values of the mass of the light doublet satisfying, $M_\pi L \ge 4$. Moreover, for each LCP 4 values of the lattice spacing were considered. This allowed us to provide the first direct lattice determination of $\ls$ with controlled continuum, finite-volume, and chiral extrapolations.

Our final result for this LEC in the SU(2) chiral limit, i.e., massless up and down quarks and physical-mass strange quark, is:
\beq
\ls \times 10^{3} = 2.79(58)_{\mathrm{stat}}(19)_{\mathrm{syst}} = 2.79(61)_{\mathrm{tot}}.
\eeq
A comparison with previous estimates is displayed in Fig.~\ref{fig:comp_l7}. Our result is compatible with, and significantly improves on, previous lattice determinations, $\ls \times 10^3 = 6.5(3.8)_{\rm stat}(0.20)_{\rm syst}$~\cite{Boyle:2015exm} (indirect determination), $\ls \times 10^3 = 2.50(1.30)_{\rm stat}(0.50)_{\rm syst}$~\cite{Frezzotti:2021ahg} (direct determination for a single gauge ensemble), as well as with the phenomenological estimate $\ls\times 10^3=7(4)$~\cite{GrillidiCortona:2015jxo}.

\acknowledgments
It is a pleasure to thank C.~Bonati for useful discussions, and M.~Gorghetto and G.~Villadoro for communications about Ref.~\cite{Gorghetto:2018ocs}. The work of C.~Bonanno is supported by the Spanish Research Agency (Agencia Estatal de Investigaci\'on) through the grant IFT Centro de Excelencia Severo Ochoa CEX2020-001007-S and, partially, by the grant PID2021-127526NB-I00, both of which are funded by MCIN/AEI/10.13039/501100011033. R.~Dionisio acknowledges funding from the European Union (EU) Next Generation EU (NGEU) -- National Recovery and Resilience Plan (NRRP) -- MISSION 4 COMPONENT 1, INVESTMENT N.4.1 -- CUP N.~I51J24000160007 (PhD Cycle XL, Ministerial Decree no.~629/2024). F.~Sanfilippo is supported by ICSC -- Centro Nazionale di Ricerca in High Performance Computing, Big Data and Quantum Computing,
funded by the EU NGEU -- and by the Italian Ministry of University and Research (MUR) project FIS 00001556. This work has also been supported by the project “Non-perturbative aspects of fundamental interactions, in the Standard Model and beyond” funded by MUR, Progetti di Ricerca di Rilevante Interesse Nazionale (PRIN), Bando 2022, grant 2022TJFCYB (CUP I53D23001440006). Numerical calculations have been performed on the \texttt{Leonardo} machine at Cineca, based on the agreement between INFN and Cineca, under projects INF23\_npqcd and INF24\_npqcd.

\appendix

\section*{Appendix}

\section{Taste symmetry restoration in the continuum limit of the pion mass}\label{app:pion_masses}

\begin{figure}[!t]
\centering
\includegraphics[scale=0.22]{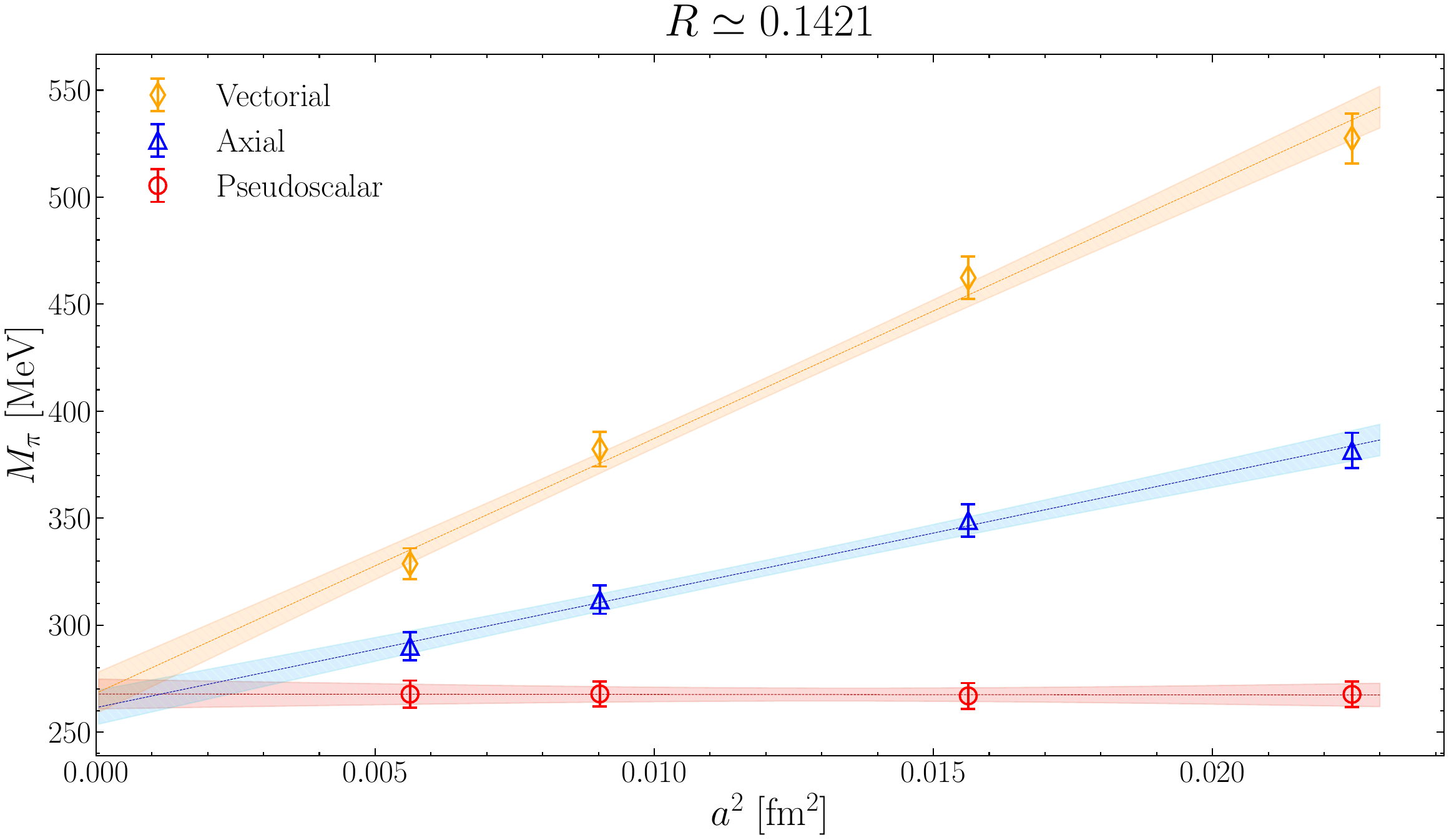}
\includegraphics[scale=0.22]{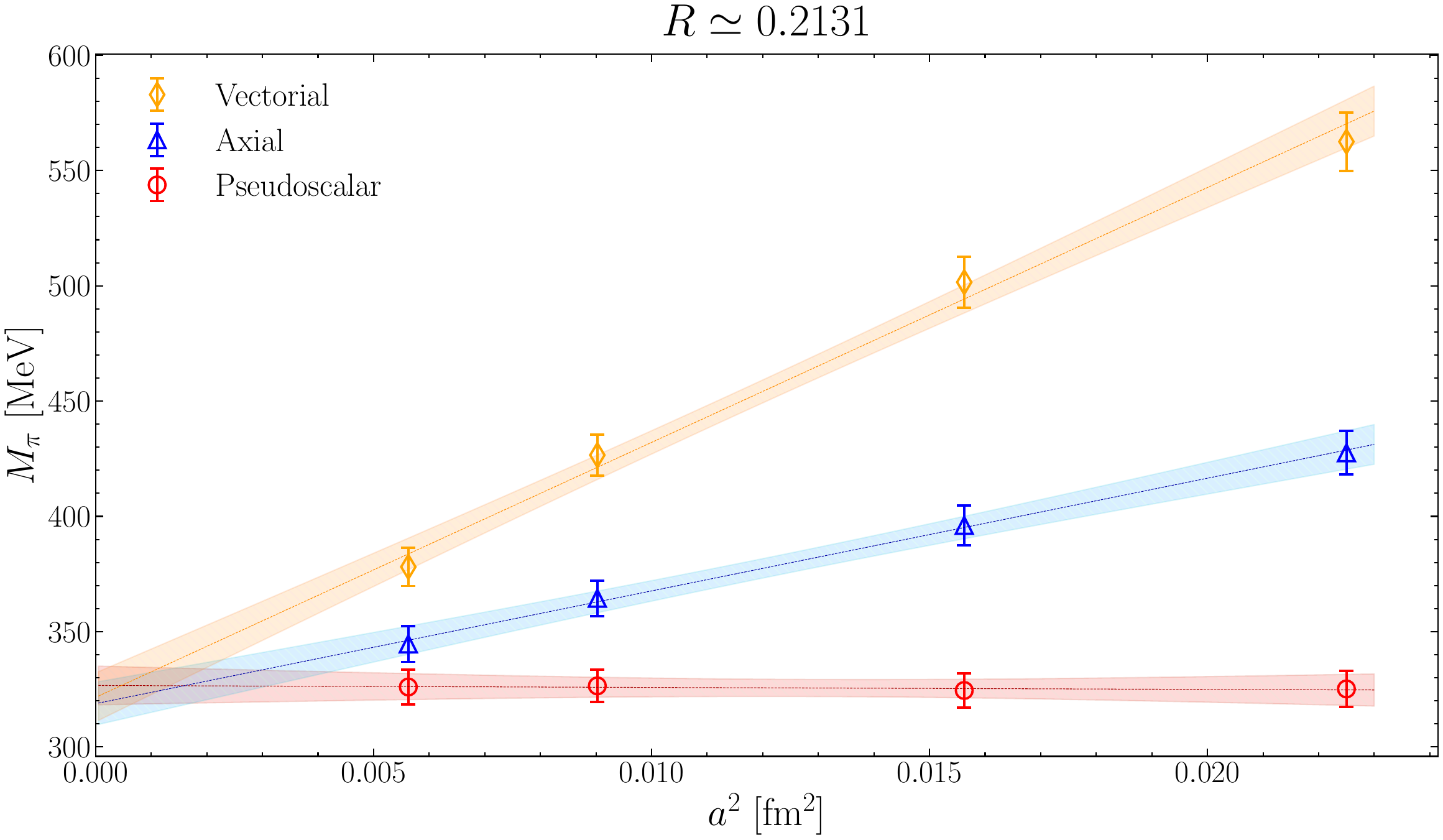}
\includegraphics[scale=0.22]{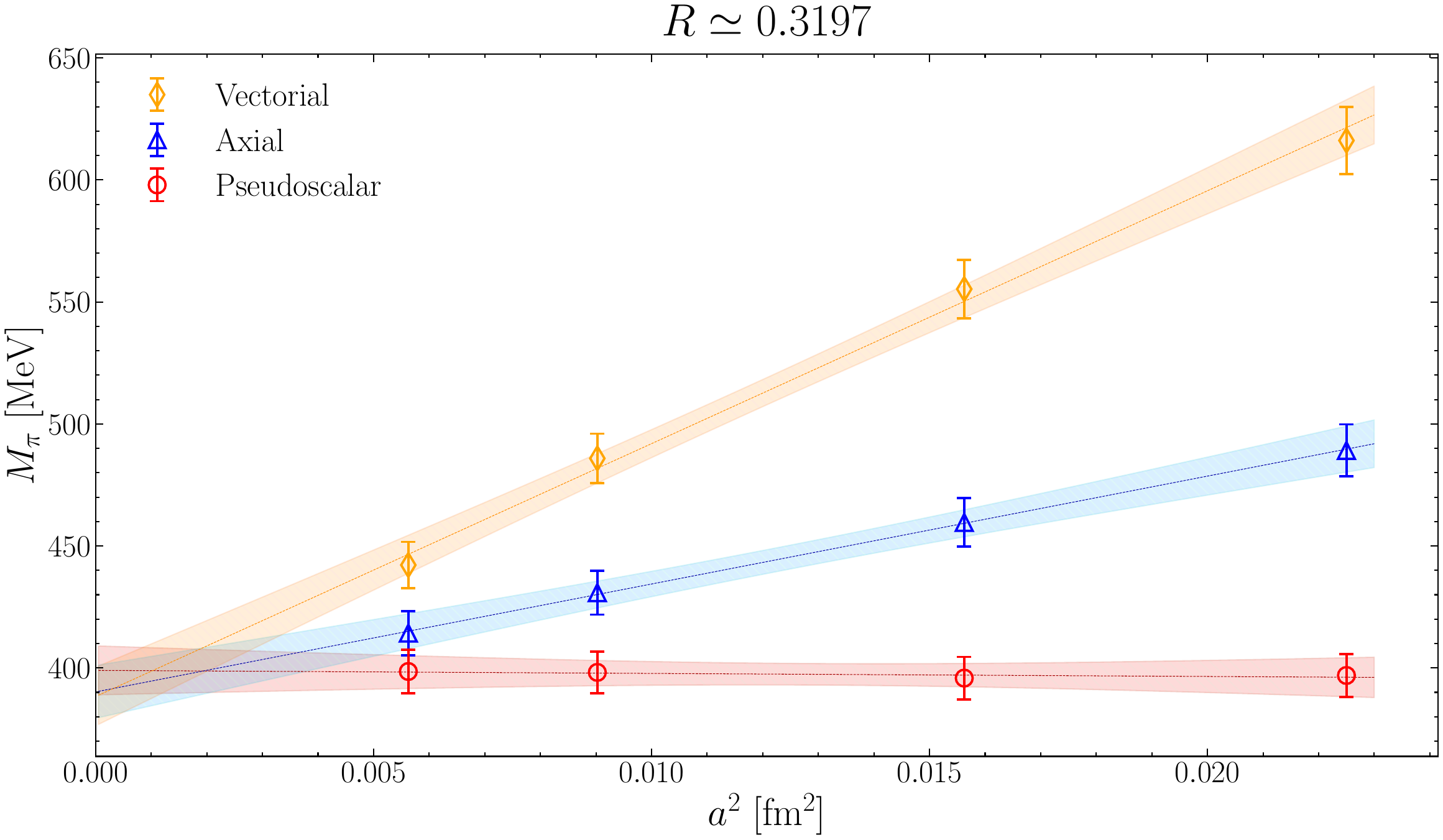}
\caption{Continuum limits of the pion masses extracted from interpolating operators with different taste structure for all the gauge ensembles employed in this study.}
\label{fig:taste_masses}
\end{figure}

The estimator in Eq.~\eqref{eq:l7_estimator} essentially involves computing the time derivative of the difference between two four-point correlators, corresponding to the connected and disconnected diagrams in Eq.~\eqref{eq:pion_mass_diff_diagrams}.
A crucial step in computing these correlators is the proper choice of interpolating operators. The relevant operators are those corresponding to the neutral and charged pion, namely $\mathcal{O}_{\pi^0}$ and $\mathcal{O}_{\pi^+}$. 

However, with staggered fermions there is no straightforward one-to-one correspondence between the pion fields in the continuum and those on the lattice. This problem arises due to the so-called taste symmetry~\cite{Borsanyi:2010cj}. Specifically, the staggered pion states are described by operators of the form $\bar{\psi} \gamma_5 \otimes \xi_{\Ssub} \psi$, where the taste matrices $\xi_{\Ssub}$ span five irreducible representations~\cite{Bernard:2007ez}:
\beq\label{eq:tasteirreps}
\xi_{\Ssub} \in \{1, \xi_5, \xi_\mu, \xi_{\mu 5}, \xi_{\mu\nu} \} = \{\text{I, P, V, A, T}\},    
\eeq
representing the singlet (I), pseudo-scalar (P), vector (V), axial-vector (A), and tensor (T) channels. Among these, the pseudo-scalar (P) channel corresponds to the pseudo-Nambu--Goldstone boson.
In the continuum limit, taste symmetry is exact, and all the staggered pions are degenerate. On the lattice, however, taste symmetry is broken at finite lattice spacing, causing the staggered pions to split into multiplets according to the tensor structure of their interpolating operators in~\eqref{eq:tasteirreps}.
An important consequence of taste symmetry breaking is that the order in which the continuum and chiral limits are taken becomes non-trivial. In the staggered formulation, these limits do not commute: to properly restore continuum physics, one must first take the continuum limit before performing the chiral extrapolation. This ensures the restoration of taste symmetry, leading to a degenerate pion spectrum.

At finite lattice spacing, however, if the chiral limit is taken first, only the mass of the pion associated with the pseudo-scalar channel vanishes. This behavior is due to the fact that staggered fermions preserve a remnant of the chiral symmetry of the continuum theory, protecting the pseudo-Nambu--Goldstone boson from acquiring a mass. Hence, a controlled continuum extrapolation is essential to recover the full taste symmetry structure and the correct physical pion spectrum~\cite{Bernard:2007ez}. As shown in Fig.~\ref{fig:taste_masses}, we have indeed verified with our data that this is the case.

\section{Comparison of \texorpdfstring{$\ls$}{l7} determinations from the pseudo-scalar, vector and axial-vector channels}\label{app:comp_channels}

\begin{figure}[!t]
\centering
\includegraphics[scale=0.255]{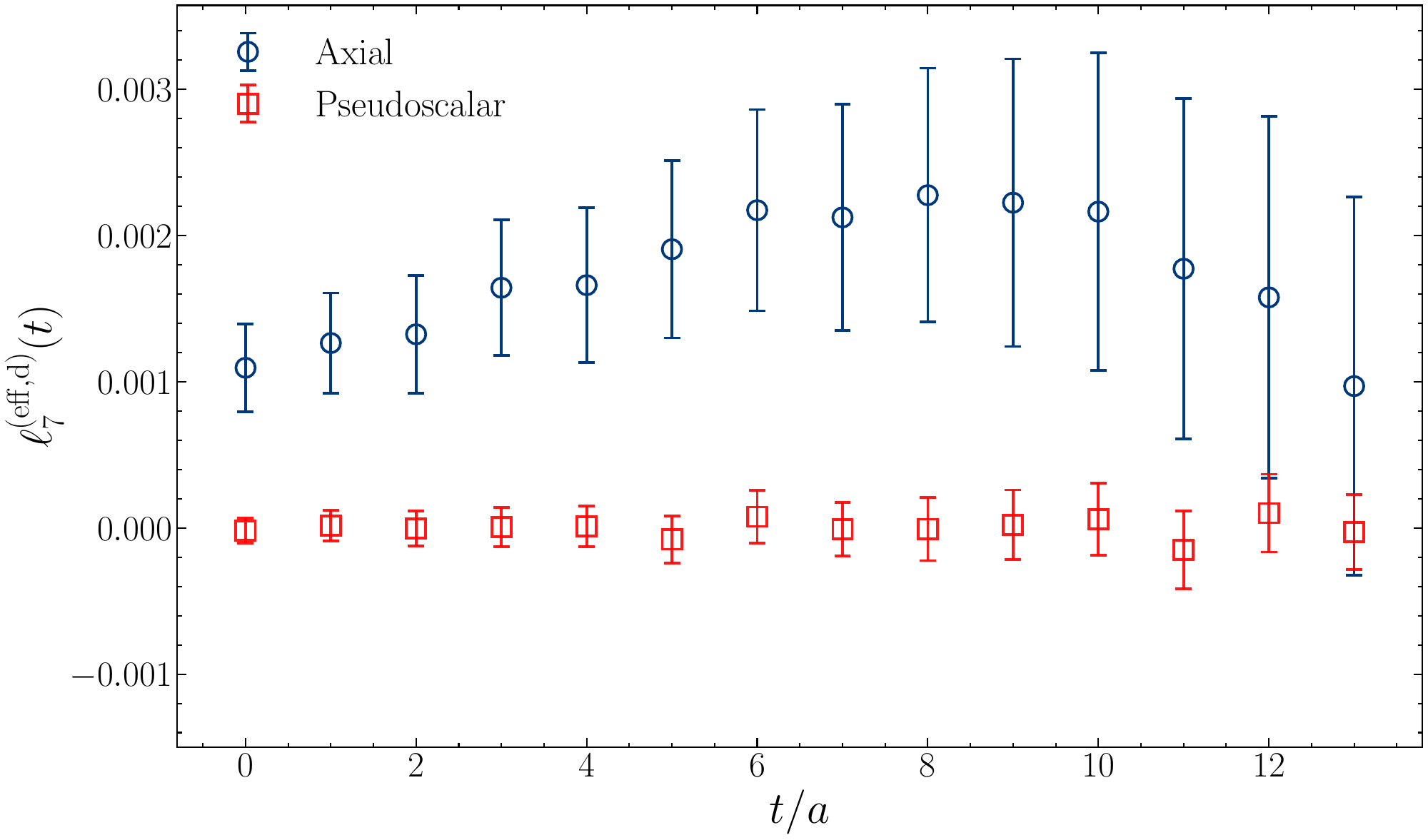}
\caption{Comparison of the disconnected contribution to $\ls$ obtained from the pseudoscalar and the axial channels, for $a \simeq 0.0964$~fm and $R \simeq 0.1421$.}
\label{fig:disc_comp}
\end{figure}

As outlined in Sec.~\ref{sec:res}, no signal was obtained for the disconnected contribution in the customary pseudoscalar channel, motivating our choice of using the axial channel throughout the calculations presented in this study. In this appendix we show, in Fig.~\ref{fig:disc_comp}, the improvement in the signal for the disconnected contribution $\ls^{\effdsup}$ obtained using the axial channel with respect to the pseudoscalar one.

\begin{figure}[!t]
\centering
\includegraphics[scale=0.22]{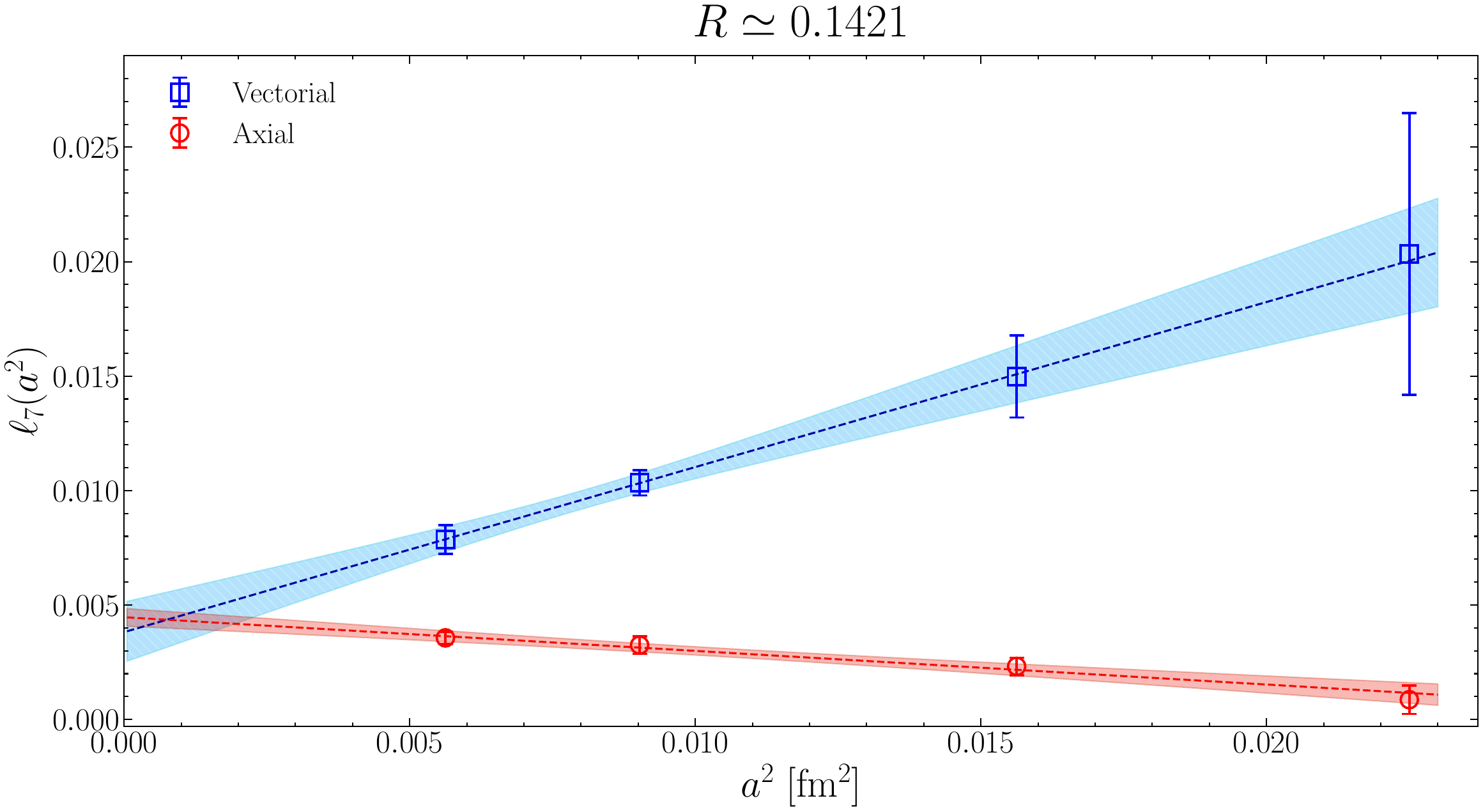}
\caption{Comparison of our determinations of $\ls$ from the vector and the axial-vector channels for $R\simeq 0.1421$.}
\label{fig:vector_vs_axial_l7}
\end{figure}

As a further check of our procedure, we also extracted $\ls$ using Eq.~\eqref{eq:l7_estimator} from the vector channel. Despite results differing from the axial-vector ones at finite lattice spacing, the two determinations nicely converge towards the same continuum limit, see Fig.~\ref{fig:vector_vs_axial_l7}. This is not surprising, giving that all interpolating operators, in the continuum limit, describe exactly the same particle with the same mass, cf.~also Appendix~\ref{app:pion_masses}. Since the axial-vector channel gave smaller uncertainties with respect to the vector one, we chose the former to perform our final calculations.

\begin{figure}[!t]
\centering
\includegraphics[scale=0.21]{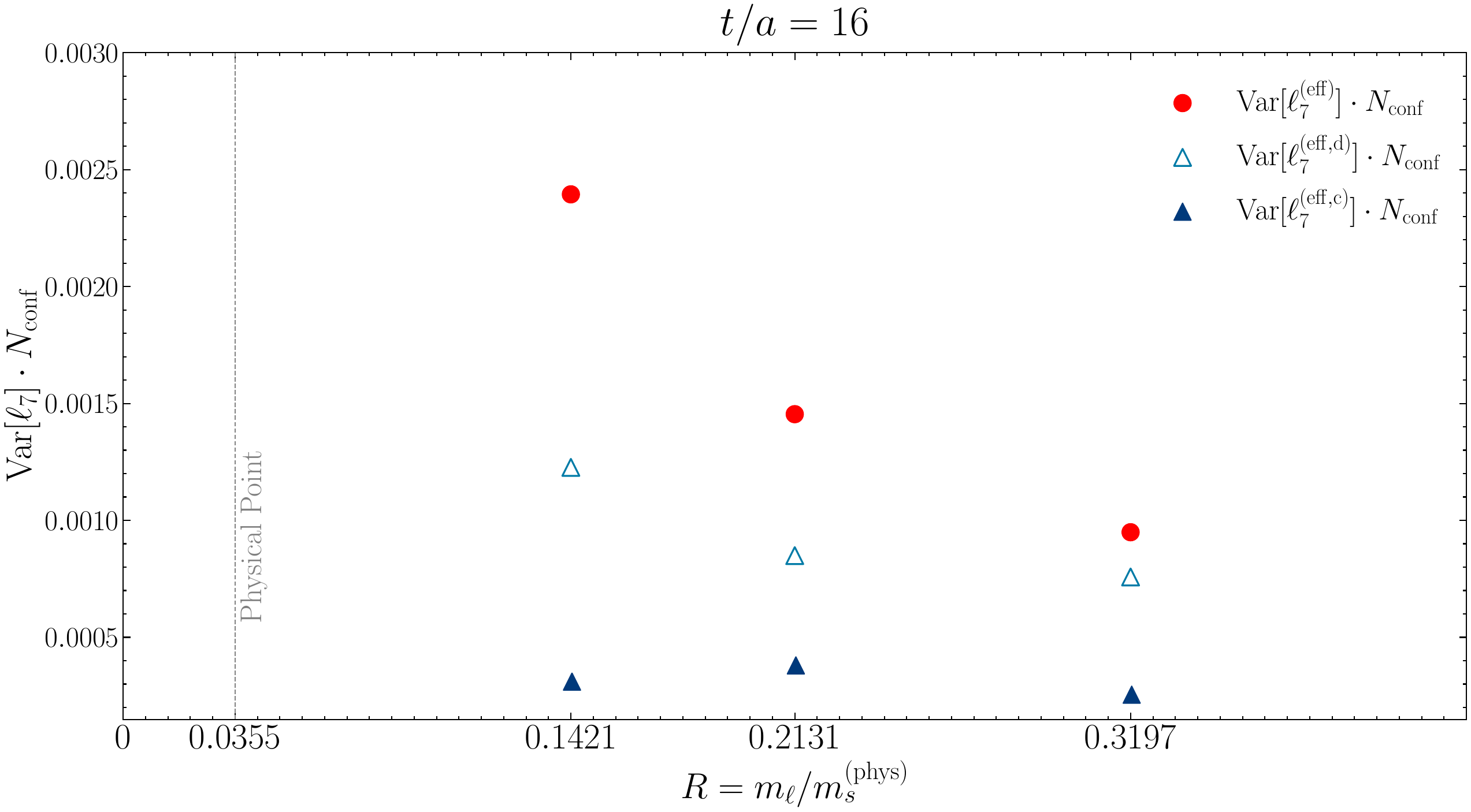}
\caption{Scaling of the rescaled variance of $\ls^{\effsup}(t)$ (and its connected/disconnected components) for a time slice in the plateau region ($t/a=16$) as a function of $R=\mql/\mqs^{\physsup}$ for the finest lattice spacing of each LCP considered in this study.}
\label{fig:error}
\end{figure}

\section{Scaling of the signal-to-noise ratio of \texorpdfstring{$\ls$}{l7} towards the chiral limit}\label{app:error_scaling}

Our final result for $\ls$ is dominated by statistical uncertainties due to the chiral extrapolation. Thus, improving our final error would require to perform direct determinations of $\ls$ for lower values of $\mql$. However, we observed a rapid deterioration of the signal-to-noise ratio of this quantity when approaching the chiral limit.

In order to quantify this effect, we studied the scaling of the statistical variance of $\ls^{\effsup}(t)$ as a function of the quark mass for a fixed time slice in the region where the plateau is observed. For each mass, we considered the finest available lattice spacing $a\simeq 0.075$ fm, and computed the variance of $\ls^{\effsup}$ --- both for the full estimator and for its connected and disconnected components --- multiplied by the total number of configurations of the corresponding ensemble. This rescaling allowed us to isolate the intrinsic noise of our observable of choice in a statistics-independent way. The results, shown in Fig.~\ref{fig:error}, display a clear rapid growth of the rescaled variance as the chiral limit is approached.

Furthermore, as shown in the main text (see Sec.~\ref{sec:res}), the signal for $\ls$ tends to decrease linearly with the quark mass towards the chiral limit. The combination of a diminishing signal and of rapidly-growing statistical fluctuations makes a reliable determination of $\ls$ at, say, the physical point, currently unfeasible with the typical statistics that can be generated with an affordable numerical effort.

\section{Charged and neutral pion masses at NNLO in \texorpdfstring{$\ChPT$}{ChiPT}}\label{app:NNLO_pion_mass_splitting}

At NNLO in the chiral expansion, the charged and neutral pion masses can be written as follows:
\beq
M^2_{\pi^+} &=& M^2 (1+m_1 + m_2),\\
\nonumber\\[-1em]
\nonumber\\[-1em]
\Delta M_\pi^2 &=& M^2 \Delta^2 (\delta_1 + \delta_2),
\eeq
with $\Delta M_\pi^2 = M_{\pi^+}^2-M_{\pi^0}^2$ the strong pion mass splitting. In these formulas, $M^2=2B\mql$, $\Delta \equiv \Delta \mql / \mql$, while $m_1$, $\delta_1$ and $m_2$, $\delta_2$ are, respectively, the NLO and NNLO $\ChPT$ corrections for the charged pion mass and the pion mass splitting. Their explicit expressions read:
\beq
\label{eq:m1}
m_1 &=& 2\frac{M^2}{F^2}\left[\ell^r_{\scriptscriptstyle{3}}(\mu) + L(\mu)\right],\\
\label{eq:m2}
m_2 &=& \frac{M^4}{F^4}\left\{ \frac{17}{8} L^2(\mu) -16 C_2(\mu)\right.\\
&& - \left[(14 \ell_1^r+8 \ell_2^r+3\ell_3^r)(\mu)+\frac{49}{192\pi^2}\right]L(\mu)\nonumber \\
&& + \frac{1}{16\pi^2} (\ell_1^r+2 \ell_2^r+l_3^r)(\mu)+\frac{163}{96}\frac{1}{(16\pi^2)^2} \nonumber \\
&&\left. - \Delta^2 \left[\ell_7 L(\mu) + 16 C_2^\prime(\mu)\right] \right\},\nonumber\\
\label{eq:delta1}
\delta_1 &=& 2\frac{M^2}{F^2} \ls ,\\
\label{eq:delta2}
\delta_2 &=& 2\frac{M^4}{F^4}\left[-3 \ls L(\mu) + 16 C_{\IBsub}(\mu) - \frac{\ls}{32\pi^2} \right].
\eeq
In Eqs.~\eqref{eq:m1}--\eqref{eq:delta2}, we have used the following short-hands:
\beq
L(\mu) \equiv \frac{1}{16\pi^2}\log\left(\frac{M^2}{\mu^2}\right).
\eeq
is the chiral log, while
\beq
C_2 &\equiv& 2 c_6^r+c_7^r+2 c_8^r+c_9^r\\
&&-3 c_{10}^r-6 c_{11}^r-2 c_{17}^r-4 c_{18}^r,
\nonumber\\
C_2^\prime &\equiv& c_7^r-c_9^r-3 c_{10}^r-2 c_{11}^r,\\
C_{\IBsub} &\equiv& c^r_9 -3c_{10} - 2 c^r_{11} -3 c^r_{17} - 2 c^r_{18} - 4 c^r_{19},
\eeq
indicate specific combinations of the 57 renormalized LECs $c_i^r$ appearing in the 2-flavor chiral Lagrangian in $\ChPT$ at NNLO.

The expressions for $m_1$ and $\delta_1$ date back to Gasser and Leutwyler~\cite{Gasser:1982ap,Gasser:1983yg,Gasser:1984gg}, while the expression for $m_2$ was previously computed in Refs.~\cite{Burgi:1996qi,Bijnens:1997vq} only in the isospin-symmetric case $\Delta=0$, and neglecting the contributions coming from the NNLO LECs, as these were only introduced later~\cite{Bijnens:1999sh,Bijnens:1999hw}. Thus, the full expression of $m_2$ (including isospin-breaking corrections and the NNLO LECs contributions) and of $\delta_2$ are new results first presented in this study. Note also that our NNLO formula for $M_{\pi^0}^2$ corrects a misprint in Eq.~(A6) of Ref.~\cite{Gorghetto:2018ocs}.\footnote{As confirmed by the authors of~\cite{Gorghetto:2018ocs} via private communication.}

Coming back to $\Delta M_\pi^2$, the running of $C_{\IBsub}(\mu)$ and $L(\mu)$ is such that $\delta_2$ is renormalization-group invariant. Thus, one may also rewrite the NNLO correction to the strong pion mass splitting as:
\beq
\delta_2 &=& -6 \ls \frac{M^4}{16\pi^2 F^4}\log\left(\frac{M^2}{\Lambda^2_{\IBsub}}\right), 
\eeq
with
\beq
\Lambda_{\IBsub} \equiv \mu \exp\left\{\frac{128\pi^2}{3}\frac{C_{\IBsub}(\mu)}{\ls}-\frac{1}{12}\right\}
\eeq
a physical (i.e., $\mu$-independent) energy scale characterizing isospin breaking at NNLO.

Adopting the representation for $\delta_2$ in Eq.~\eqref{eq:delta2}, setting
\beq
\mu=\mu_0 \equiv 4\pi F = 1.088(8)\text{ GeV},
\eeq
and substituting
\beq
C\equiv 16 \frac{C_{\IBsub}(\mu_0)}{F^2}-\frac{\ls}{32\pi^2F^2},
\eeq
one easily obtains Eq.~\eqref{eq:pion_mass_splitting_NNLO} in the main text. From the chiral fit of the corrected data $\tilde{\ell}_{\scriptscriptstyle{7}}(R)$, we instead extracted the quantity $\tilde{C}$, cf.~Eq.~\ref{eq:chiral_fit_final}, which is related to $C_{\IBsub}(\mu_0)$ via the following relation:
\beq
\tilde{C} = 2 \frac{B \mqs}{F^2} \left[16C_{\IBsub}(\mu_0)-\frac{\ls}{32\pi^2}\right].
\eeq
Since we found, from the chiral fit,
\beq
{\tilde{C} = (-1.4\pm 2.4)\cdot 10^{-3},}
\eeq
we obtain:
\beq
{C_{\IBsub}(\mu_0) = (-8 \pm 24)\cdot 10^{-7}.}
\eeq
This value for $C_{\IBsub}(\mu_0)$ implies:
\beq
\frac{\Lambda_{\IBsub}}{\mu_0} &=& \frac{\Lambda_{\IBsub}}{4\pi F} = 0.87(32),
\eeq
\beq
\implies \Lambda_{\IBsub} &\simeq& 0.95(35) \text{ GeV}.
\eeq

\section{Determination of \texorpdfstring{$\lt$}{l3}  and \texorpdfstring{$\lf$}{l4} }\label{app:l3_l4}

\begin{figure}[!t]
\centering
\includegraphics[scale=0.31]{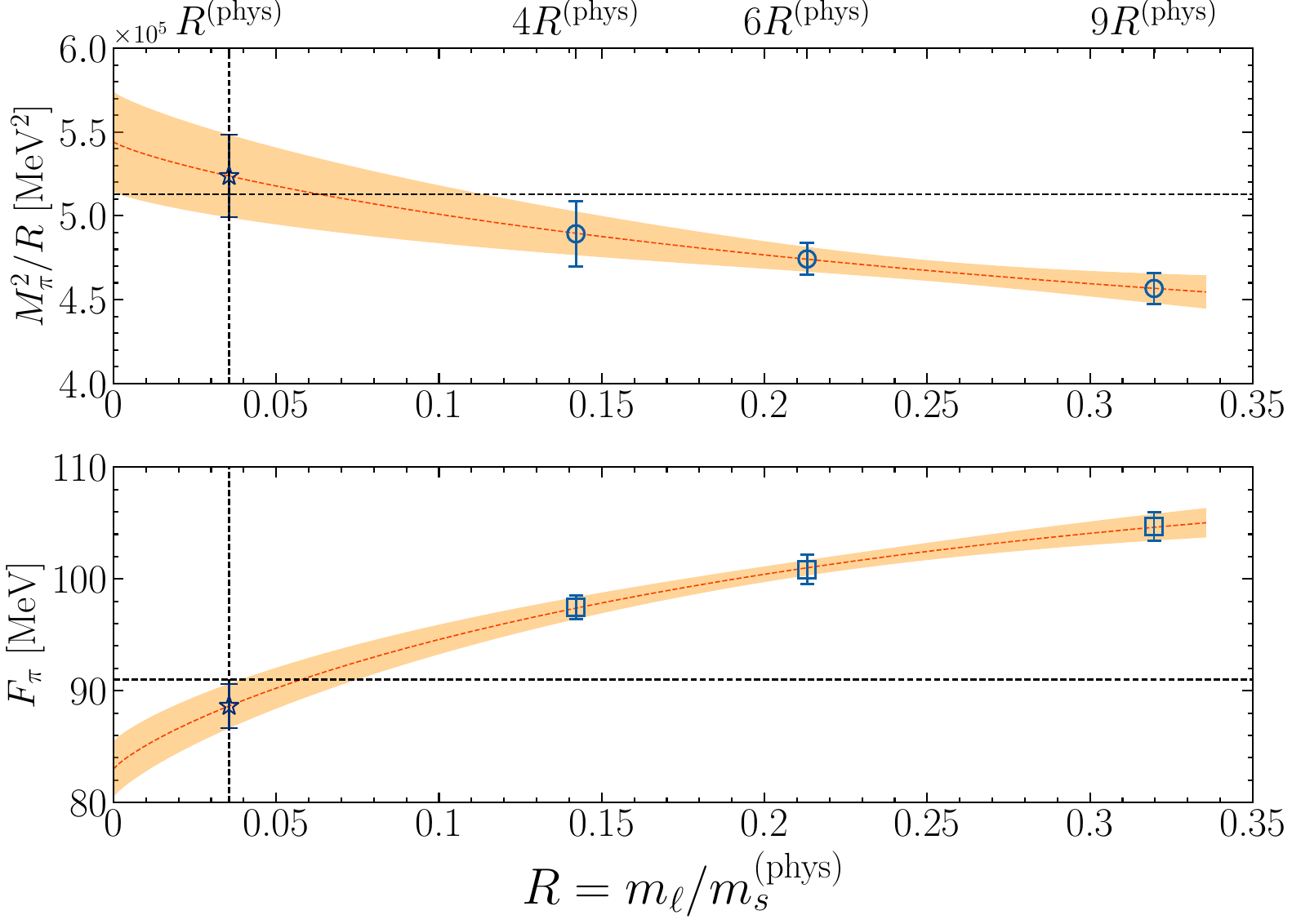}
\caption{Joint chiral fit of the determinations of $M_\pi(R)$ and $F_\pi(R)$ of Tab.~\ref{tab:lattice_params} as a function of the quark mass ratio $R=\mql/\mqs^{\physsup}$ according to NLO $\chi$PT predictions in Eqs.~\eqref{eq:fit_func_l3} and~\eqref{eq:fit_func_l4}. Dashed lines represent the physical point $R\simeq 0.036$~\cite{FlavourLatticeAveragingGroupFLAG:2021npn} and the physical values of $M_\pi$ and $F_\pi$. Starred points represent our extrapolations in $R=R^{\physsup}$.}
\label{fig:chiral_fit_l3_l4}
\end{figure}

\begin{figure}[!t]
\centering
\includegraphics[scale=0.32]{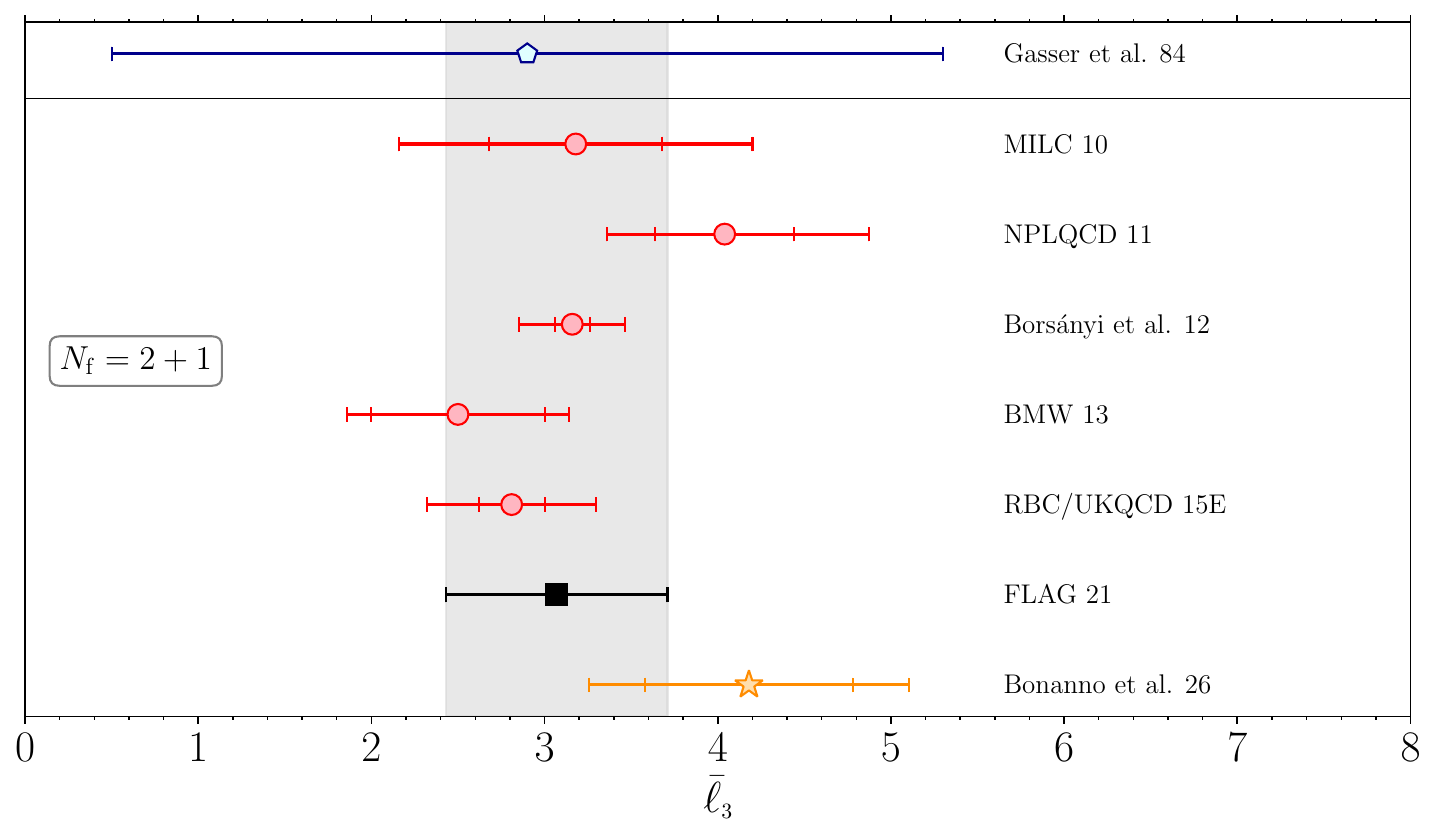}
\includegraphics[scale=0.32]{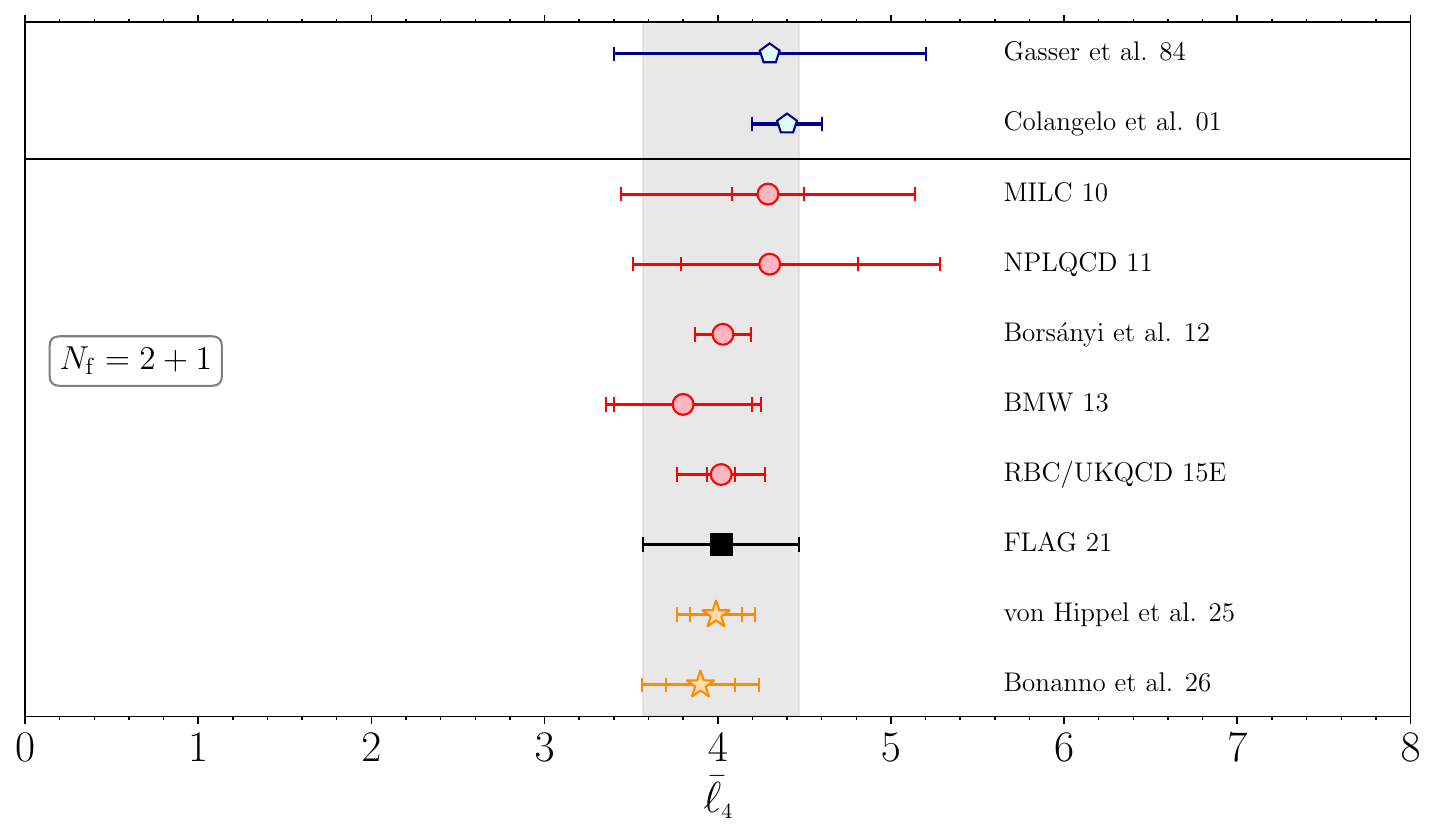}
\caption{Comparison of our $N_{\rm f}=2+1$ lattice determinations of $\lt$ (top panel) and $\lf$ (bottom panel), drawn as starred points, with previous ones~\cite{MILC:2010hzw,Beane:2011zm,Borsanyi:2012zv,BMW:2013fzj,Boyle:2015exm} entering FLAG 21 averages~\cite{FlavourLatticeAveragingGroupFLAG:2021npn} (square points and shaded bands). For $\lf$ we also reported the recent determination of~\cite{vonHippel:2025uhr,Ottnad:2025hkw}. We also show the phenomenological estimates of~\cite{Gasser:1983yg,Colangelo:2001df}.}
\label{fig:comp_l3_l4}
\end{figure}

The determinations of $M_\pi$ and $F_\pi$ reported in the present study in Tab.~\ref{tab:final}, and used for the calculation of $\ls$, can also be employed to extract two other NLO LECs: $\lt$ and $\lf$. These couplings parameterize the corrections to the chiral limits of the pion mass and of the pion decay constant~\cite{Gasser:1984gg}:
\beq
M_\pi^2 &=& M^2 \left\{1- \frac{M^2}{32\pi^2F^2} \left[\lt - L(M)\right]\right\},\\
\nonumber\\
\nonumber\\
F_\pi &=& F \left\{1 + \frac{M^2}{16\pi^2F^2} \left[\lf - L(M)\right]\right\},
\eeq
with
\beq
L(M) &=& \log\left(\frac{M^2}{{M_\pi^{\physsup}}^2}\right),\\
\nonumber\\
M_\pi^{\physsup} &=& 135\text{ MeV},\\
\nonumber\\
\nonumber
\eeq
a chiral log. Using $M^2=2B\mql = 2 B \mqs R$, one can perform a four-parameter global fit of $M_\pi(R)$ and $F_\pi(R)$ and extract $\lt$ and $\lf$:
\beq
\label{eq:fit_func_l3}
\frac{M_\pi^2(R)}{R} &=& 2B\mqs \left\{1- \frac{B\mqs R}{16\pi^2F^2} \left[\lt - \tilde{L}(R)\right]\right\},\\
\nonumber\\
\label{eq:fit_func_l4}
F_\pi(R) &=& F \left\{1 + \frac{B\mqs R}{8\pi^2F^2} \left[ \lf - \tilde{L}(R) \right]\right\},\\\nonumber
\eeq
where now the chiral log reads
\beq
\tilde{L}(R) = \log\left(\frac{2B\mqs R}{{M_\pi^{\physsup}}^2}\right).
\eeq
The joint fit was performed after the application of the finite-volume corrections $R_{M_\pi}$ and $R_{F_\pi}$ (cf.~Sec.~\ref{sec:FSE}), which turned out to be of always smaller than our statistical uncertainties.

The LO LECs $\Sigma \mqs = B\mqs \times F^2$ and $F$ were already given in the original paper~\cite{Bonanno:2023xkg}, while $\lt$ and $\lf$ are reported here for the first time. We find:
\beq
\lt &=& 4.18(60)_{\rm stat}(70)_{\rm syst} = 4.18(92)_{\rm tot},\\
\nonumber\\
\lf &=& 3.90(20)_{\rm stat}(27)_{\rm syst} = 3.90(34)_{\rm tot}.
\eeq
The analysis for $\lt$ and $\lf$ goes along the same lines of the one for $\ls$ presented in the main text. The central values and the statistical uncertainties are obtained from the joint chiral fit including all available points, shown in Fig.~\ref{fig:chiral_fit_l3_l4}. The systematic error is the result of the quadrature sum of two error sources. One is the propagation of the systematic uncertainties on $F_\pi$ and $M_\pi$ due to the continuum extrapolation. The other is the observed difference in the fit results for $\lt$ and $\lf$ when including/excluding the point at the heaviest pion mass from the joint chiral fit.

Our chiral extrapolations of $F_\pi$ and $M_\pi$ are well under control, as it can be seen by the very good agreement of our results in $R=R^{\physsup}\simeq 0.036$~\cite{FlavourLatticeAveragingGroupFLAG:2021npn} with the physical values $M_\pi^{\physsup}=135$ MeV and $F_\pi^{\physsup}=92$ MeV. Moreover, our results for $\lt$ and $\lf$ are in very good agreement both with the latest FLAG world averages~\cite{FlavourLatticeAveragingGroupFLAG:2021npn} and with previous determinations in the literature. This comparison is illustrated in Fig.~\ref{fig:comp_l3_l4}.

\end{document}